\documentclass[usenames,dvipsnames,submission, Phys]{SciPost}

\usepackage{graphicx}
\usepackage{amssymb}
\usepackage{amsmath}
\usepackage{amsfonts}
\usepackage{bm}
\usepackage{dsfont}
\usepackage{multirow}

\usepackage{hyperref}
\hypersetup{
   unicode=true,          
   colorlinks=true,       
   linkcolor=violet,          
   citecolor=blue,        
   urlcolor=blue           
}

\graphicspath{{.}{./EPS/}}

\newcommand* {\vek}[1]{{\ensuremath{\bm{\mathrm{#1}}}}}
\newcommand* {\kk}{\vek{k}}

\newcommand* {\ee}{\ensuremath{\mathrm{e}}}
\newcommand*{\goo}[1]{\textcolor{OliveGreen}{#1}}
\newcommand*{\foo}[1]{\textcolor{magenta}{#1}}

\begin{document}

\begin{center}{\Large \textbf{
Accurate projective two-band description of topological superfluidity in
spin-orbit-coupled Fermi gases
}}\end{center}

\begin{center}
J. Brand\textsuperscript{1},
L.~A. Toikka\textsuperscript{1,2},
U. Z\"ulicke\textsuperscript{3*}
\end{center}

\begin{center}
{\bf 1} Dodd-Walls Centre for Photonic and Quantum Technologies,
Centre for Theoretical Chemistry and Physics, and New Zealand
Institute for Advanced Study, Massey University, Private Bag 102904
NSMC, Auckland 0745, New Zealand
\\
{\bf 2} Center for Theoretical Physics of Complex Systems, Institute for
Basic Science (IBS), Daejeon 34051, Republic of Korea
\\
{\bf 3} Dodd-Walls Centre for Photonic and Quantum Technologies,
School of Chemical and Physical Sciences, Victoria University of
Wellington, Wellington 6140, New Zealand
\\
* uli.zuelicke@vuw.ac.nz
\end{center}

\begin{center}
\today
\end{center}


\section*{Abstract}
{\bf
The interplay of spin-orbit coupling and Zeeman splitting in ultracold
Fermi gases gives rise to a topological superfluid phase in two spatial
dimensions that can host exotic Majorana excitations. Theoretical
models have so far been based on a four-band Bogoliubov-de Gennes
formalism for the combined spin-$\mathbf{1/2}$ and particle-hole
degrees of freedom. Here we present a simpler, yet accurate, two-band
description based on a well-controlled projection technique that provides
a new platform for exploring analogies with chiral \textit{p}-wave
superfluidity and detailed future studies of spatially non-uniform
situations.
}

\vspace{10pt}
\noindent\rule{\textwidth}{1pt}
\tableofcontents\thispagestyle{fancy}
\noindent\rule{\textwidth}{1pt}
\vspace{10pt}

\section{Introduction}
\label{sec:intro}

Topological superfluids and superconductors~\cite{Mizushima2016,
Sato2017} are the focus of great current interest because of their ability
to host unconventional Majorana excitations~\cite{Beenakker2016}. An
attractive route towards realization that was suggested early
on~\cite{Fu2008,Zhang2008,Sau2010,Alicea2010,Sato2010} utilizes
two-dimensional (2D) \textit{s}-wave superfluids with spin-orbit coupling.
The transition from the non-topological superfluid phase to the
topological superfluid phase in these systems is driven by increasing the
Zeeman energy splitting between spin-$\uparrow$ and
spin-$\downarrow$ single-particle states beyond the critical value where
the excitation gap for spin-$\downarrow$ particles closes. The resulting
effectively spinless superfluid state is expected to show all the hallmarks
associated with chiral \textit{p}-wave pairing~\cite{Kallin2016}, including
exotic Majorana states in vortex cores~\cite{Read2000,Ivanov2001,
Gurarie2007}. Promising experimental efforts are currently undertaken in
condensed-matter systems~\cite{Veldhorst2012,Hart2014,Pribiag2015}
and ultracold-atom gases~\cite{Huang2016,Meng2016}, which have the
potential to provide complementary insight and crucial ingredients for
topological quantum information devices~\cite{DasSarma2015}. One of
the important technical differences between these two platforms is the
way how the Zeeman spin splitting is introduced. For superconductors,
the required magnetic-field strengths are typically deleterious to
superconductivity, motivating a search for alternative
approaches~\cite{Loder2015,Wu2016,Isaev2017}. Being unencumbered
by this drawback, ultracold-atom realizations may offer a more direct
avenue towards implementation of topological superfluidity. Our present
study is intended to provide a useful tool for investigating topological
effects in superfluid spin-orbit-coupled Fermi gases and ultimately
enable the design and optimization of proof-of-concept devices.

\begin{figure}[t]
\centerline{
\includegraphics[width=0.45\textwidth]{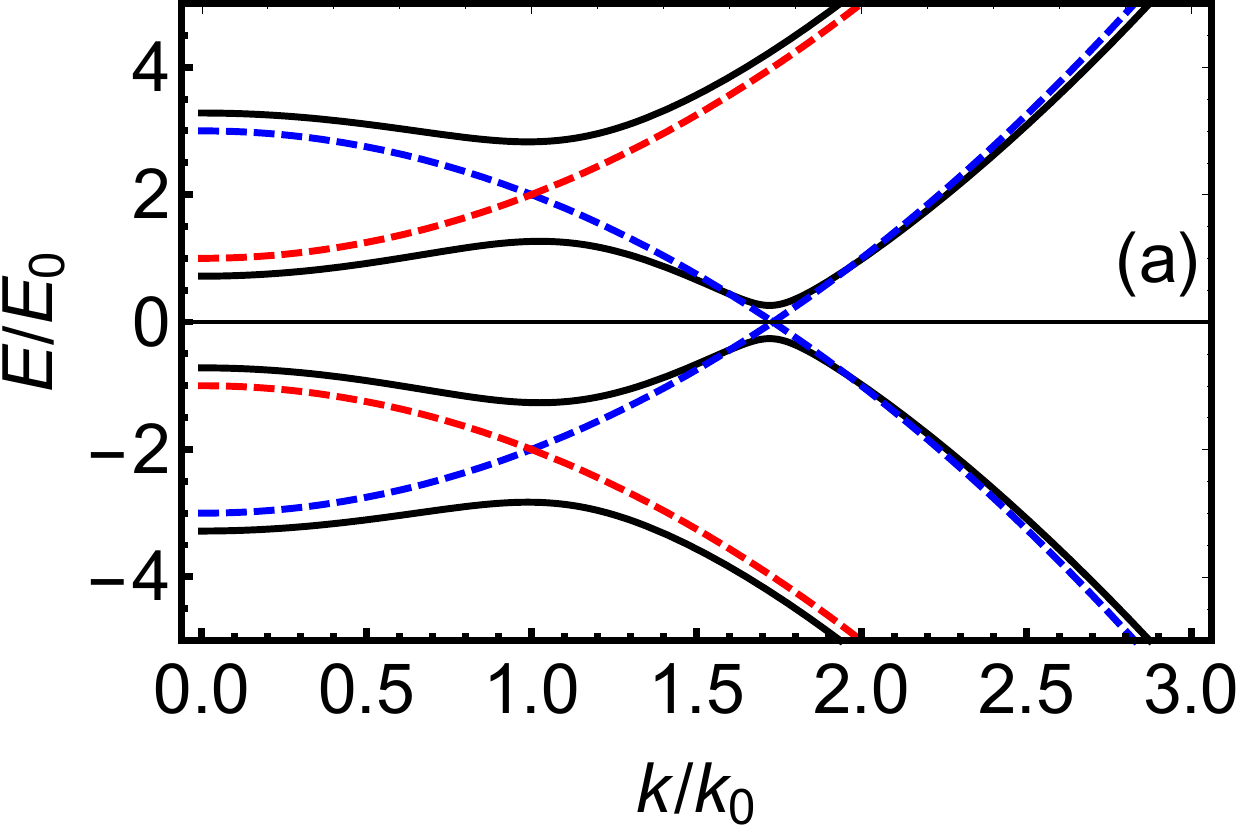}\hspace{0.5cm}
\includegraphics[width=0.45\textwidth]{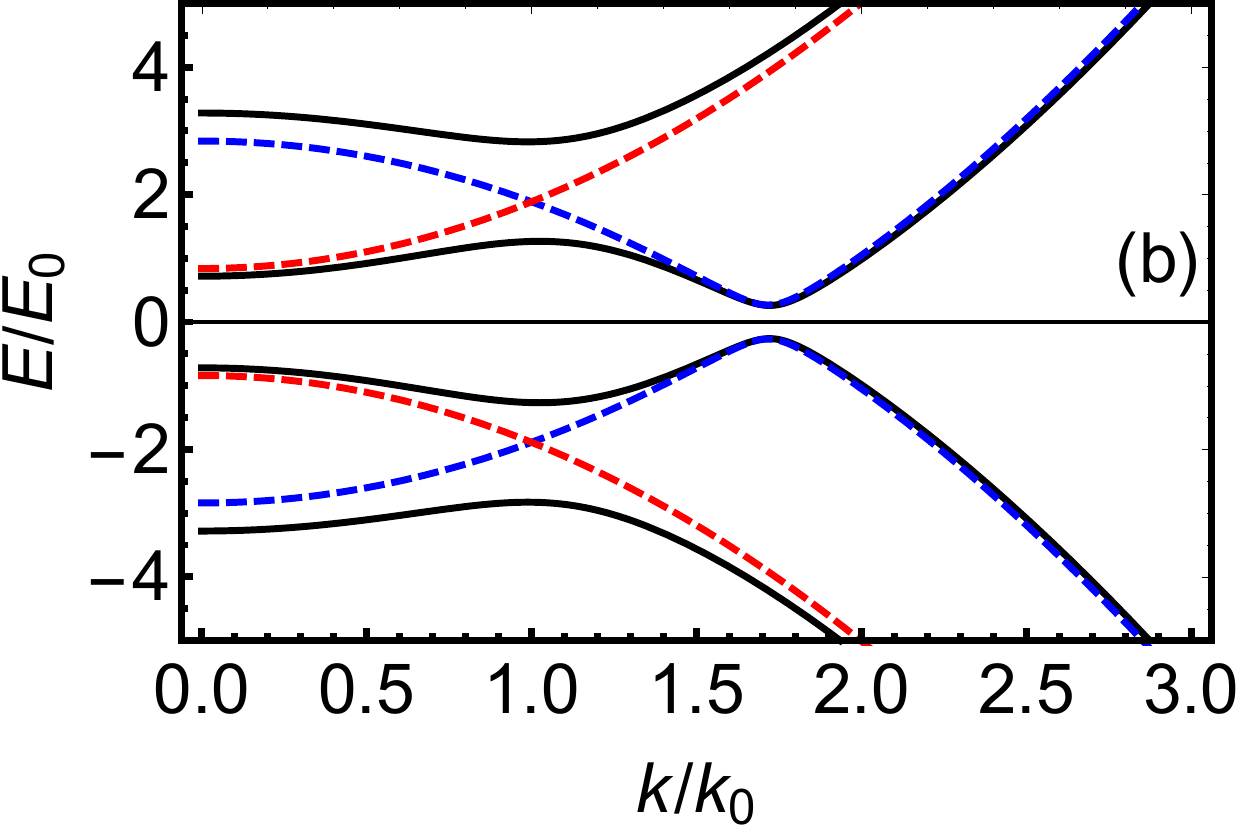}
}%
\caption{\label{fig:dispersions}%
Spectrum of Bogoliubov-quasiparticle excitation energies $E$ as a
function of wave-vector magnitude $k\equiv \sqrt{k_x^2 + k_y^2}$ in the
uniform topological-superfluid phase of a 2D Fermi gas with spin-orbit
coupling $\lambda_\kk \equiv \lambda (k_x - i\, k_y)$ and Zeeman
splitting $h$. In panel (a), the dashed blue (red) curves are the
spin-$\uparrow$ (spin-$\downarrow$) dispersions in the absence of
\textit{s}-wave pairing ($\Delta=0$) and spin-orbit coupling ($\lambda =
0$) but with large Zeeman splitting $h=2 E_0$. A finite $\Delta$ couples
spin-$\uparrow$ and spin-$\downarrow$ states, resulting in gaps
opening at $E = \pm h$ and $k = \sqrt{2 m \mu/\hbar^2}$, where $\mu$
denotes the chemical potential. In the situation depicted here, $\mu =
E_0$ with $E_0\equiv\hbar^2 k_0^2/(2 m)$ being an arbitrary energy
scale. When $\lambda$ is also finite, a third gap opens at $E=0$, and
the system is a topological superfluid for $h > \sqrt{\mu^2 + |\Delta|^2}$.
The black solid curves are the exact dispersions for $2 m\lambda/(
\hbar^2 k_0) = 0.4$ and $|\Delta|/E_0 = 0.8$. Panel (b) again depicts
these exact dispersions as black solid curves, together with the
approximate dispersions obtained by us using a Feshbach projection
onto the spin-$\uparrow$ (spin-$\downarrow$) subspace shown as the
dashed blue (red) curves.}
\end{figure}

Previous theoretical studies~\cite{Kubasiak2010,Iskin2011,Gong2011,
Yi2011,Jiang2011,Zhou2011,Tewari2011,Yang2012a,Liu2012b,
Seo2012,Yang2012,He2012a,He2013} of \textit{s}-wave pairing in
Fermi gases with spin-orbit coupling and Zeeman splitting have
examined the physical properties of these paradigmatic systems using
a mean-field treatment in a four-dimensional Nambu space. While the
breaking of spin-rotational invariance generally requires such a more
complicated~\cite{Ketterson1999}, and in general only numerically
accessible, treatment, the subspace of the spin-$\uparrow$ degrees
of freedom that are relevant for topological properties is only
two-dimensional. See Fig.~\ref{fig:dispersions}(a) for an illustration.
Thus it would be desirable to have an effective description based on
projecting into the spin-$\uparrow$ subspace. However, to discuss
manipulations of the system by controllable physical parameters and
to make predictions for experimentally accessible observables, the
influence of the spin-$\downarrow$ degrees of freedom cannot be
ignored. Here we present a fully self-consistent effective theory that
is based on an application of the Feshbach-partitioning
technique~\cite{Feshbach1958,Feshbach1962} to the spin-resolved
Bogoliubov-de~Gennes (BdG) Hamiltonian~\cite{deGennes1989,
Ketterson1999} under the assumption that the $s$-wave pair potential
$|\Delta|$ is small compared to the Zeeman energy shift $h$ that favors
spin-$\uparrow$ over spin-$\downarrow$ configurations. Our effective
theory is designed to reproduce salient features of the
Bogoliubov-quasiparticle spectrum [Fig.~\ref{fig:dispersions}(b)] and all
relevant parametric dependences associated with the topological phase
transition. This formalism is therefore ideally suited to be a platform for
further studies of topological effects, including those associated with
non-uniform superfluid phases~\cite{Grosfeld2011,Zou2016,Smith2016,
Shitade2015,Goertzen2017}. In order to be specific, and also because it
is the physically most interesting case, we develop the theory for a 2D
superfluid, but the formalism lends itself to be easily applied to 1D or 3D
situations as well. 

One of the main benefits associated with having a
spin-$\uparrow$-projected effective theory is that it facilitates the
numerical treatment of inhomogeneous and time-dependent situations.
For example, in the time-dependent study of soliton or vortex dynamics
(e.g., similar to recent work reported in Ref.~\cite{Zou2016}), the
reduction of numerical complexity obtained by moving from the original
four-spinor formalism to the projected two-spinor approach is significant.
But already for the homogeneous superfluid, the projected-theory results
are very useful because, e.g., they provide simpler expressions for the
wave functions of the Bogoliubov quasi-particle excitations and thus
facilitate convenient analytical approximations. As an example, we
derive a simple analytic expression for the chemical potential of the
spin-orbit-coupled two-dimensional superfluid [see Eq.~\eqref{eq:Mu}].

The results presented below can be compared with, and also extend,
those of previous studies of superfluidity in Fermi gases with spin-orbit
coupling and Zeeman splitting where self-consistency implied fixing the
total particle density~\cite{Iskin2011,Gong2011,Yi2011,Jiang2011,
Seo2012,Liu2012b,Yang2012,He2012a,He2013}. (In contrast,
Refs.~\cite{Zhou2011,Tewari2011,Yang2012a} consider the situation
with fixed chemical potential. See also early work~\cite{Kubasiak2010}
that focused on a lattice realization.) Most relevant benchmarking for our
present context is provided by previous works pertaining to uniform 2D
systems~\cite{Yang2012,He2012a,He2013}, but there are also useful
connections to be made with known results for trapped 2D
systems~\cite{Liu2012b} and 3D systems with 2D Rashba-type
spin-orbit coupling~\cite{Iskin2011,Gong2011,Yi2011,Jiang2011,
Seo2012}. In a slight variation on our situation of interest,
Ref.~\cite{Yang2012} considers the population imbalance between
spin-$\uparrow$ and spin-$\downarrow$ particles as a control parameter
rather than the Zeeman energy.

This article is organised as follows. In the following
Section~\ref{sec:partition}, we introduce the microscopic model for
superfluid 2D Fermi gases with spin-orbit coupling and Zeeman splitting
and apply Feshbach partitioning to derive effective theories describing
the spin-$\uparrow$ and spin-$\downarrow$ sectors separately. The
obtained formalism is applied in Section~\ref{sec:selfCon} to devise a
fully self-contained procedure for finding the chemical potential $\mu$
and \textit{s}-wave pair potential $\Delta$ for uniform systems at fixed
total particle density $n\equiv n_\uparrow + n_\downarrow$ entirely
within the $2\times 2$-projected theory for the spin-$\uparrow$ sector.
The efficacy of this approach is demonstrated in
Section~\ref{sec:TheoComp} by presenting a comparison of predictions
for phase boundaries and thermodynamic quantities obtained within the
effective two-band and exact four-band theories. Following the usual
convention, we measure relevant parameters in units of the
density-defined magnitude of the 2D Fermi wave vector $k_\mathrm{F}
= \sqrt{2\pi n}$ and associated Fermi energy $E_\mathrm{F} = \hbar^2
k_\mathrm{F}^2/(2 m) \equiv \pi\hbar^2 n/m$, with $m$ denoting the
single-particle mass. Our conclusions and an outlook toward future work
are presented in the final Section~\ref{sec:concl}.

\section{Feshbach partitioning of the BdG Hamiltonian}
\label{sec:partition}

Our starting point is the Bogoliubov-de~Gennes (BdG) equation
describing quasiparticle excitations in a superfluid Fermi gas without
spin-rotational invariance. It reads~\cite{deGennes1989,Ketterson1999}
\begin{subequations}
\begin{align}\label{eq:bdg}
\mathcal{H} \left(\begin{array}{c} u^\uparrow \\ u^\downarrow \\
v^\uparrow \\ v^\downarrow \end{array}\right)  = E \left(
\begin{array}{c} u^\uparrow \\ u^\downarrow \\ v^\uparrow \\
v^\downarrow \end{array}\right) \quad ,
\end{align}
with complex spinor entries $u^\sigma$ ($v^\sigma$) denoting quantum
amplitudes of spin-$\sigma$ particle (hole) states in a Bogoliubov
excitation of the superfluid. (Here and in the following, $\sigma\in
\{\uparrow,\downarrow\}$ is used as a compact label for the spin degree
of freedom.) The Hamiltonian matrix in four-dimensional particle-hole
(Nambu) space is
\begin{align}\label{eq:origBdG}
\mathcal{H} = \left( \begin{array}{cccc} \epsilon_{\kk \uparrow}
-\mu &\lambda_\kk & 0& -\Delta \\ \lambda_\kk^* & \epsilon_{\kk
\downarrow} - \mu &\Delta & 0 \\ 0 & \Delta^* & -\epsilon_{\kk
\uparrow} + \mu & {\lambda_\kk^*} \\ -\Delta^* & 0 &\lambda_\kk &
-\epsilon_{\kk\downarrow} +\mu \end{array} \right) \quad ,
\end{align}
\end{subequations}
where `$\ast$' indicates complex conjugation, $\kk \equiv (k_x, k_y)$ is
the 2D wave vector, and $\epsilon_{\kk\uparrow (\downarrow)}=
\epsilon_\kk \, \substack{-\\(+)} \, h$ with $\epsilon_\kk = \hbar^2 (k_x^2
+k_y^2)/2m$. For spatially inhomogeneous configurations, $k_j \equiv
-i \partial_j$ is to be treated as an operator while, for a homogeneous
superfluid, it can be replaced by its wave-number eigenvalue. The 
Zeeman energy splitting is denoted by $h$, and $\lambda_\kk$ is the
spin-orbit coupling. Examples of typically considered $\kk$-linear
spin-orbit couplings are the 2D-Dirac~\cite{Winkler2015},
2D-Rashba~\cite{Bychkov1984,Winkler2003}, and
2D-Dresselhaus~\cite{Eppenga1988,Winkler2003} types that
correspond to different functional forms $\lambda_\kk = \lambda (k_x -
i k_y)$, $\lambda\, i(k_x - i k_y)$, and $\lambda (k_x + i k_y)$,
respectively, but are all unitarily equivalent. In particular, the eigenvalue
spectrum of $\mathcal{H}$ for the homogeneous superfluid depends on
the spin-orbit coupling only via the quantity $|\lambda_\kk|^2 \equiv
\lambda^2 (k_x^2+k_y^2)$ and therefore has the same functional form
for all three of the above-mentioned spin-orbit-coupling types. The
eigenvalue spectrum $E_{\kk\alpha, \eta}$ is characterised by four
bands of dispersion relations as shown in Fig.~\ref{fig:dispersions}(a) for
a particular set of parameters. In order to be able to refer to a specific
band, we introduce the following naming convention (for the fully gapped
case), where $\alpha=+(-)$ indicates states that
have energy $E\ge 0$ ($E\le 0$), whereas the index $\eta =\,\, >$
($<$) labels the higher(lower)-energy pair of excitation branches; i.e.,
$|E_{\kk\alpha,>}| \ge |E_{\kk\alpha, <}|$. For what follows, it will be
useful to know the asymptotic behavior of the dispersions in the limit of
large 2D-wave-vector magnitude $k\equiv \sqrt{k_x^2 + k_y^2}$. We find
\begin{equation}\label{eq:asympDisp}
E_{\kk+,<(>)} = \epsilon_\kk - \mu \, \substack{-\\(+)} \, \sqrt{h^2 +
|\lambda_\kk|^2} + \mathcal{O} \left( \frac{|\Delta|^2}{\epsilon_\kk}\right)
\quad .
\end{equation}

The matrix equation (\ref{eq:bdg}) can be reorganized by forming
$2\times 2$ sub-blocks on the diagonal that are associated with
subspaces for fixed spin degree of freedom,
\begin{subequations}
\begin{equation}\label{eq:F1}
\left(\begin{array}{cc} {\mathcal H}^{\uparrow\uparrow} & {\mathcal
H}^{\uparrow\downarrow} \\ {\mathcal H}^{\downarrow\uparrow} &
{\mathcal H}^{\downarrow\downarrow} \end{array}\right) \left( 
\begin{array}{c} w^\uparrow \\ w^\downarrow \end{array} \right) = E \left(
\begin{array}{c} w^\uparrow \\ w^\downarrow \end{array}\right) \quad ,
\end{equation}
with the definitions
\begin{align}
w^\sigma =& \left(\begin{array}{c} u^\sigma \\ v^\sigma \end{array}
\right) \quad ,\\[0.1cm]
{\mathcal H}^{\sigma\sigma} =& \left(\begin{array}{cc} \epsilon_{\kk
\sigma} - \mu & 0 \\ 0 &  -\epsilon_{\kk\sigma} + \mu \end{array}\right)
\quad ,\\[0.1cm]
{\mathcal H}^{\uparrow\downarrow} \equiv& \, \big( {\mathcal
H}^{\downarrow\uparrow} \big)^\dagger = \left( \begin{array}{cc}
\lambda_\kk & -\Delta \\ \Delta^* & \lambda_\kk^* \end{array}\right)
\quad , \label{eq:offdiag}
\end{align}
\end{subequations}
and `$\dagger$' denoting Hermitian conjugation. Simple algebra yields
$2\times 2$-matrix equations that formally decouple the individual spin
sectors,
\begin{subequations}\label{eq:Feshbach}
\begin{align}\label{eq:y}
w^{\bar\sigma} =& - \left(  {\mathcal H}^{\bar\sigma\bar\sigma} - E\,
\mathds{1} \right)^{-1} {\mathcal H}^{\bar\sigma\sigma}\, w^\sigma
\quad , \\ \label{eq:proj}
\left[ {\mathcal H}^{\sigma\sigma} - {\mathcal H}^{\sigma\bar\sigma}
\left(  {\mathcal H}^{\bar\sigma\bar\sigma} - E\, \mathds{1} \right)^{-1}
{\mathcal H}^{\bar\sigma\sigma} \right] w^\sigma = & \, E\, w^\sigma
\quad , 
\end{align}
\end{subequations}
where $\mathds{1}$ is the $2\times 2$ identity matrix and $\bar\sigma$
denotes the opposite of $\sigma$; i.e., $\bar\sigma=\,\downarrow
(\uparrow)$ if $\sigma=\,\uparrow(\downarrow)$. While formally exact
and a $2\times 2$ BdG-like equation in spin-$\sigma$ space,
Eq.~\eqref{eq:proj} is not really useful without approximations, since the
unknown energy eigenvalue $E$ appears on both sides of the equation.
Assuming $\Delta$ to be small compared to other relevant energy
scales, we replace $E$ in the denominator by the exact $\Delta=0$
solution from  Eq.~(\ref{eq:asympDisp}) to obtain the approximation
\begin{subequations}\label{eq:approx}
\begin{align}
\left( {\mathcal H}^{\downarrow\downarrow} - E\, \mathds{1} \right)^{-1}
\approx & \,\, \frac{1}{2 h_\kk}\left( \begin{array}{cc} 1 & 0 \\ 0 & -1
\end{array}\right) \quad , \\[0.1cm]
\left( {\mathcal H}^{\uparrow\uparrow} - E\, \mathds{1} \right)^{-1}
\approx & \,\, \frac{1}{2 h_\kk}\left( \begin{array}{cc} -1 & 0 \\ 0 & 1
\end{array}\right) \quad .
\end{align}
\end{subequations}
Here $2 h_\kk = h + \sqrt{h^2 +|\lambda_\kk|^2}$ can be thought of as a
$k$-dependent effective Zeeman splitting that is modified by the
presence of the spin-orbit coupling. The approximation becomes good if
$2 h_\kk$ is large compared to the neglected term, i.e.~when $|\Delta|^2
\ll \epsilon_{\kk}h_\kk$, which is always asymptotically true for large
$k$. Substituting the approximations (\ref{eq:approx}) into \eqref{eq:proj}
yields truly decoupled BdG equations for the individual spin sectors,
\begin{equation}\label{eq:2bandBdG}
{\mathcal H}^\sigma_\mathrm{eff} \, w^\sigma = E^\sigma\, w^\sigma
\quad ,
\end{equation}
where
\begin{subequations}
\begin{align} \label{eq:BdGapprox}
{\mathcal H}^\sigma_\mathrm{eff} =& \left(\begin{array}{cc} 
\xi_{\kk\sigma} & {\tilde\Delta}_{\kk\sigma} \\[0.2cm]
{\tilde \Delta}^\ast_{\kk\sigma} & -\xi_{\kk\sigma} \end{array}\right) 
\quad , \\[0.2cm]
\xi_{\kk\uparrow(\downarrow)} =& \,\, \epsilon_{\kk\uparrow(\downarrow)}
\,\, \substack{+\\(-)} \,\, \frac{|\Delta|^2 - |\lambda_\kk|^2}{2 h_\kk} - \mu
\quad , \\[0.2cm]
{\tilde\Delta}_{\kk\uparrow(\downarrow)} =&\,\, - \left\{
\lambda_\kk^{(\ast)} \, ,\, \frac{\Delta}{h_\kk} \right\} \quad .
\end{align}
\end{subequations}
We adopted an anticommutator notation $\{ A\, , B\} = (A B + B A)/2$
to be able to incorporate situations with spatially inhomogeneous
\textit{s}-wave pair potential\footnote{Spatial inhomogeneity will also
require a suitable treatment of the $\kk$-dependence in $h_\kk$.}
$\Delta$. Notice that the off-diagonal matrix element in
(\ref{eq:BdGapprox}) that is responsible for opening the gap in the
quasiparticle spectrum is given by $\lambda_\kk\, \Delta/h_\kk$, which
is proportional to both the spin-orbit-coupling strength $\lambda$ and
$\Delta$. In the case of $\kk$-linear spin-orbit coupling and uniform
$\Delta$, its leading-order dependence on $\kk$ resembles the pair
potential for a chiral-\textit{p}-wave superfluid. The emergence of
\textit{p}-wave pairing in the present context was inferred in earlier
works~\cite{Zhang2008,Alicea2010,Sato2010,Seo2012} through a
transformation into the so-called `helicity' basis of the single-particle
Hamiltonian in the presence of spin-orbit coupling. Although instructive
at the time, this transformation does not provide a useful basis for
in-depth quantitative studies. In contrast, our present work enables a
precise derivation of the effective pairing potential of spin-$\sigma$
states, including systematic corrections to the chiral-\textit{p}-wave form.

Assuming a spatially uniform pair potential $\Delta$, straightforward
diagonalization of (\ref{eq:BdGapprox}) yields approximate energy
dispersions and corresponding eigenspinors for the spin-$\sigma$
subspace as
\begin{subequations}
\begin{eqnarray}\label{eq:projDisp}
E^\sigma_{\kk\alpha} &=& \alpha\,\sqrt{\xi_{\kk\sigma}^2 +
\frac{|\lambda_\kk|^2 |\Delta|^2}{h_\kk^2}} \quad , \\[0.4cm]
\label{eq:projSpinor} w^{\downarrow(\uparrow)}_{\kk\alpha} &=&
\sqrt{N^{\downarrow(\uparrow)}_{\kk\alpha}} \left( \begin{array}{c}
\sqrt{\frac{E^{\downarrow(\uparrow)}_{\kk\alpha} +\, \xi_{\kk\downarrow
(\uparrow)}}{2 E^{\downarrow(\uparrow)}_{\kk\alpha}}} \\[0.4cm] - \alpha
\, \frac{\lambda_\kk^{(*)}}{|\lambda_\kk|}\, \frac{\Delta^*}{|\Delta|} \,
\sqrt{\frac{E^{\downarrow(\uparrow)}_{\kk\alpha} -\, \xi_{\kk\downarrow
(\uparrow)}}{2 E^{\downarrow(\uparrow)}_{\kk\alpha}}}\end{array} \right)
\,\, . \quad
\end{eqnarray}
\end{subequations}
The $N^\sigma_{\kk\alpha}$ are normalization factors that have to be
found  from the four-spinor normalization condition $1 = \sum_\sigma
\big( w^\sigma\big)^\dag w^\sigma$. Using the substitution (\ref{eq:y}),
this condition translates into separate normalization conditions for the
individual spin sectors,
\begin{subequations}
\begin{align}
\label{eqn:normX}
1 &= \big(w^\sigma_{\kk\alpha}\big)^\dagger\, \mathcal{L}_{\kk
\alpha}^\sigma \, w^\sigma_{\kk\alpha} \quad ,\\[0.2cm]\label{eqn:normL}
\mathcal{L}_{\kk\alpha}^\sigma &= \mathds{1} + {\mathcal H}^{\sigma
\bar\sigma} \left( {\mathcal H}^{\bar\sigma\bar\sigma} - E^\sigma_{\kk
\alpha}\, \mathds{1} \right)^{-2} {\mathcal H}^{\bar\sigma\sigma} \quad .
\end{align}
\end{subequations}
Further application of the approximations (\ref{eq:approx}) in
(\ref{eqn:normL}) yields
\begin{equation}\label{eq:normUp}
N^\sigma_{\kk\alpha} \approx \left[ 1 + \frac{|\Delta|^2 +
|\lambda_\kk|^2}{4 h_\kk^2} \right]^{-1} \quad .
\end{equation}

\begin{figure}[t]
\includegraphics[width=0.32\textwidth]{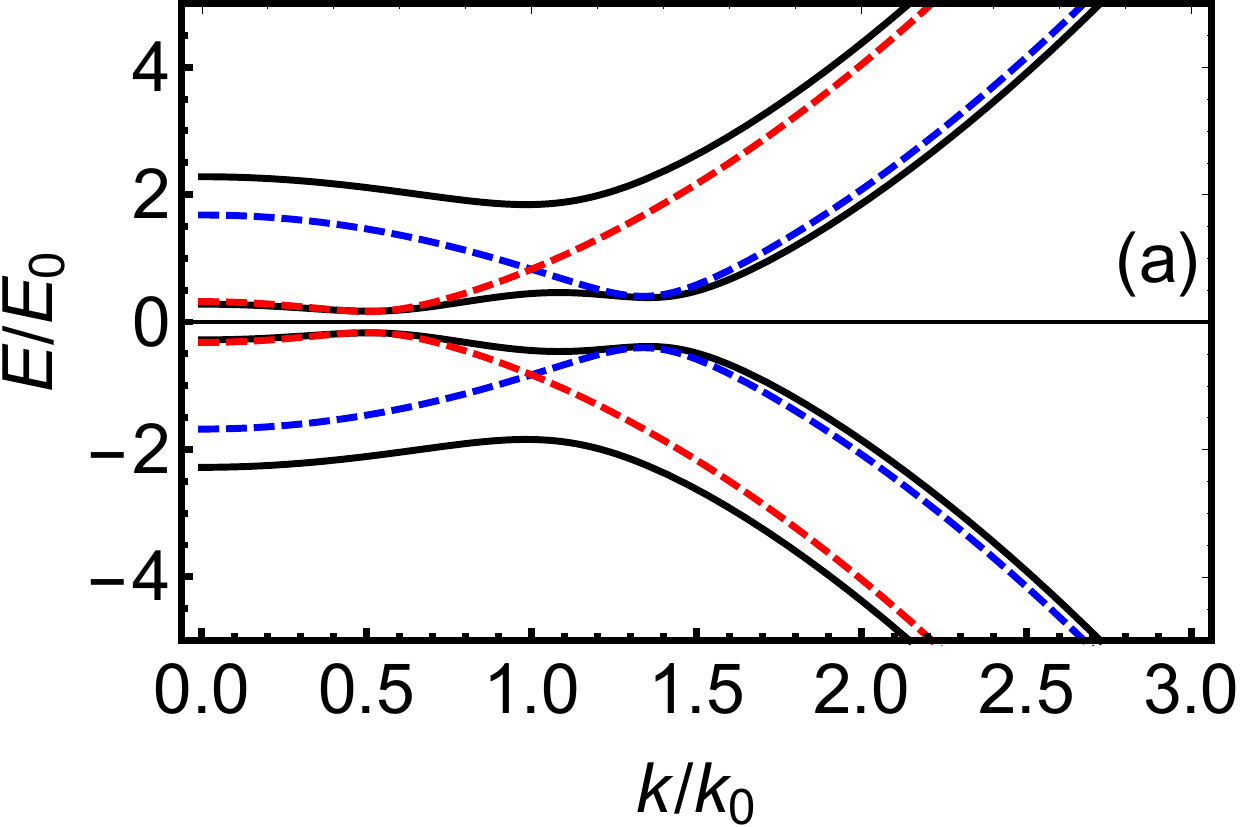}\hfill
\includegraphics[width=0.32\textwidth]{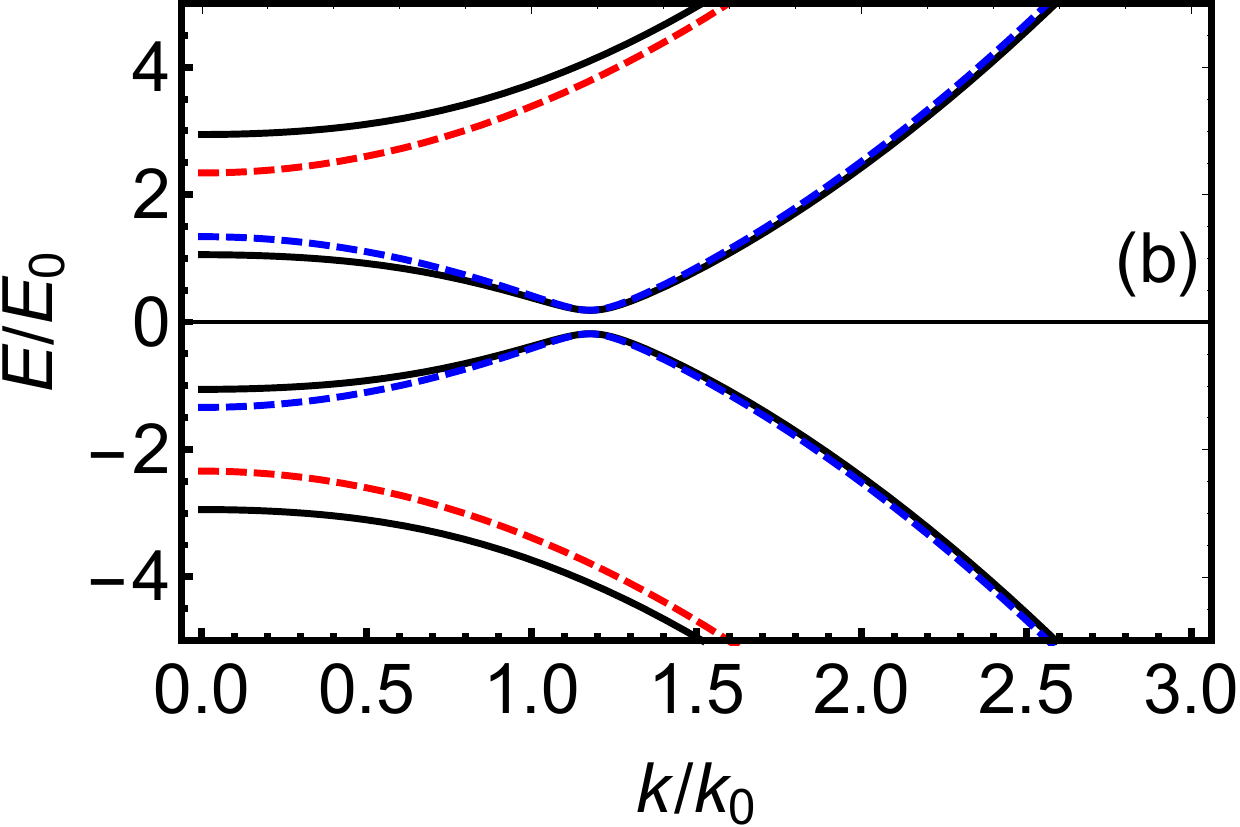}\hfill
\includegraphics[width=0.32\textwidth]{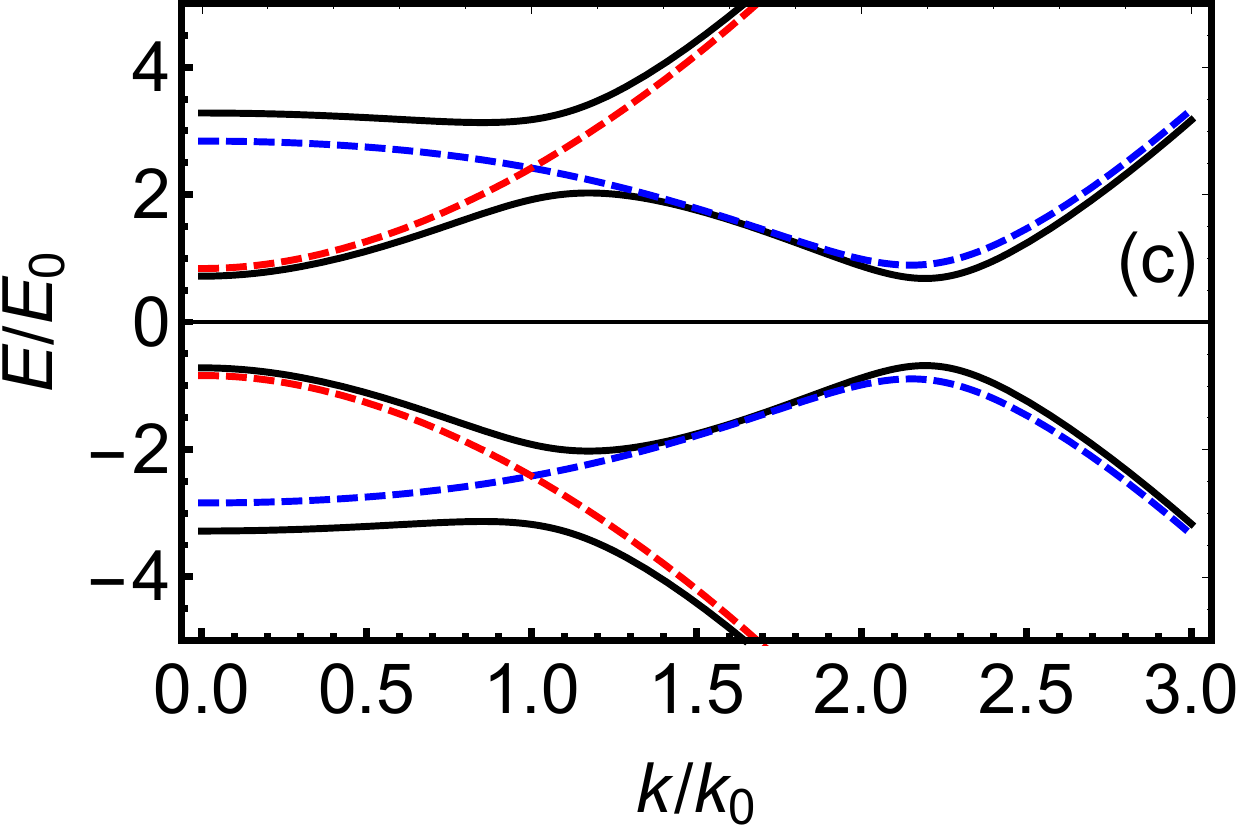}
\caption{\label{fig:moreDisp}%
More comparisons between the exact Bogoliubov-quasiparticle
dispersions for the $4\times 4$ BdG Hamiltonian (\ref{eq:origBdG})
(black solid curves) with the approximate $2\times 2$-projection results
$E_{\kk\alpha}^\sigma$ from (\ref{eq:projDisp}) [dashed blue (red)
curve for $\sigma=\,\uparrow(\downarrow)$]. Panel~(a) shows an
example for the situation where the superfluid is non-topological [$h =
E_0$, $2 m\lambda/(\hbar^2 k_0) = 0.4$, $\mu= E_0$, and $|\Delta|/E_0
= 0.8$]. The case depicted in Panel~(b) is for a topological superfluid
having a negative chemical potential, which typically occurs for large
two-particle binding energies and/or large spin-orbit-coupling strengths
[here $h = 2 E_0$, $2 m \lambda/(\hbar^2 k_0) = 0.4$, $\mu= -0.5\,
E_0$, and $|\Delta|/E_0 = 0.8$]. Panel~(c) illustrates deviations
occurring when spin-orbit coupling becomes quite large [$h = 2 E_0$,
$2 m \lambda/(\hbar^2 k_0) = 1.5$, $\mu= E_0$, and $|\Delta|/E_0 =
0.8$].}
\end{figure}

Figure~\ref{fig:dispersions}(b) shows a comparison between the exact
dispersions $E_{\kk\alpha,\eta}$, obtained by diagonalizing the original
$4\times 4$ BdG Hamiltonian (\ref{eq:origBdG}), and the approximate
results $E_{\kk\alpha}^\sigma$ from (\ref{eq:projDisp}), calculated using
the $2\times 2$-subspace projections. While the projected theory does
not reproduce the \textit{s}-wave pairing gaps around $E = \pm h$ and
$k=\sqrt{2 m \mu/\hbar^2}$, it describes very well the region around the
topological gap at $E=0$. Most crucially, as it turns out, the dispersions
for large $k$ are correctly given by the effective $2\times 2$-projected
approach. As shown in the following, this enables a faithful description of
relevant thermodynamic properties. Further comparisons between the
exact $4\times 4$-theory dispersions and the $2\times 2$-projection
approximations are explored in Fig.~\ref{fig:moreDisp}. The low-energy
gap structure turns out to be well-described even in the non-topological
regime, as illustrated in Fig.~\ref{fig:moreDisp}(a). Overall good
agreement is achieved in situations when the chemical is
negative\footnote{Instances where the chemical potential of a
superfluid becomes negative include the BEC regime of the
BCS-BEC crossover~\cite{Randeria1990,Parish2015}, and systems with
large spin-orbit coupling~\cite{He2012}.} due to the absence of any
crossing points between opposite-spin dispersions
[Fig.~\ref{fig:moreDisp}(b)]. The topological gap ceases to be
well-described for quite large spin-orbit-coupling strengths
[Fig.~\ref{fig:moreDisp}(c)].

Our Feshbach-projection approach embodied in
Eqs.~\eqref{eq:Feshbach} and \eqref{eq:approx} differs from common
perturbative methods such as the Schrieffer-Wolf
transformation\footnote{See, e.g., Appendix~B in
Ref.~\cite{Winkler2003} for a detailed discussion, or the original
work~\cite{Schrieffer1966}.} in two crucial aspects. Firstly, the
dispersions \eqref{eq:projDisp} derived from our $2\times 2$-projected
Bogoliubov-de~Gennes Hamiltonians \eqref{eq:BdGapprox} are
well-behaved at all $\kk$, whereas those obtained, e.g., from the
Schrieffer-Wolf transformation become singular at the $s$-wave pairing
gap because of a degeneracy between the eigenvalues of ${\mathcal
H}^{\uparrow\uparrow}$ and ${\mathcal H}^{\downarrow\downarrow}$ at
this point. Secondly, by careful choice of the approximation
\eqref{eq:approx}, we are able to reproduce the large-$|\kk|$
asymptotics of the exact energy dispersions within the $2\times
2$-projected approach, which cannot be achieved by perturbation theory
because it treats the spin-sector couplings given in
Eq.~\eqref{eq:offdiag} as small quantities.

By construction, the projected-theory results for quasiparticle dispersions
and wave functions become strictly exact in the limit of vanishing
$s$-wave pair potential $\Delta$, i.e., when no avoided crossings occur.
Thus, for finite $\Delta$, we may expect all quantities that do not
explicitly depend on the avoided crossing between the spin-$\uparrow$
and spin-$\downarrow$ dispersions to be described correctly as long as
$|\Delta|\ll h$.

\section{Self-consistency for uniform systems with fixed density}
\label{sec:selfCon}

Knowledge of the Bogoliubov-quasiparticle excitations permits
calculation of all physical quantities of interest~\cite{deGennes1989,
Ketterson1999}. For simplicity, we focus on the zero-temperature limit in
the following. Generalization to the case of finite temperatures is
straightforward~\cite{deGennes1989,Ketterson1999} but does not add
any crucial insights for our present purpose.

The pair potential $\Delta$ can be expressed in terms of the
eigenspinors of the BdG equation (\ref{eq:bdg}) and the strength $g$ of
attractive interactions in the \textit{s}-wave channel as
\begin{equation}\label{eq:DeltaSC}
\Delta = -\frac{g}{2\Omega} \sum_{\kk,\eta} \left[ u^\uparrow_{\kk - ,\eta}
\big(v^\downarrow_{\kk - ,\eta} \big)^* + u^\downarrow_{\kk + ,\eta}
\big( v^\uparrow_{\kk + ,\eta} \big)^* \right] \quad ,
\end{equation}
where $\Omega$ denotes the system volume (here: area). As the
quasiparticle excitation energies and associated spinor amplitudes are
themselves functions of $\Delta$, Eq.~(\ref{eq:DeltaSC}) constitutes a
self-consistency condition~\cite{deGennes1989,Ketterson1999}.
However, the expression (\ref{eq:DeltaSC}) is formally divergent and
needs to be regularized using the relation~\cite{Randeria1990}
\begin{equation}
\label{eqn:4x4renorm}
\frac{1}{g} = -\frac{1}{\Omega}\sum_\kk \frac{1}{ 2\epsilon_\kk +
E_{\mathrm b}}\quad ,
\end{equation}
where $E_{\mathrm b}>0$ is the absolute value of the binding energy
of the two-particle bound state in 2D~\cite{Petrov2000,Petrov2001,
Petrov2002} in the absence of spin-orbit coupling, i.e., for $\lambda=
0$. (Modifications of the two-body bound state in a quasi-2D Fermi gas
due to spin-orbit coupling are discussed in Refs.~\cite{Zhou2012,
He2012,Zhang2012}, but these are not relevant for the pairing-gap
regularisation procedure~\cite{Zhai2015}.) The binding energy is related
to the 2D scattering length via the expression $E_\mathrm{b} = 4\ee^{-2
\gamma}\hbar^2/(m a_\mathrm{2D}^2)$, where $\gamma = 0.577\dots$
is the Euler constant\footnote{Generally, $E_{\mathrm b} \sim\hbar^2/
(m a_\mathrm{2D}^2)$ for shallow dimers, but values given in the
literature for the prefactor on the r.h.s.\ of that relation vary. This is due to
different conventions being used when defining the two-dimensional
scattering length $a_\mathrm{2D}$~\cite{Liu2010a,Bertaina2011,
Levinsen2015}. Our choice follows related previous work~\cite{He2012a,
He2013}.}. Recent experimental realizations of low-temperature 2D
Fermi gases~\cite{Fenech2016,Boettcher2016,Hueck2017} have been
able to access a wide parameter range $-7\lesssim \ln( k_\mathrm{F}
a_\mathrm{2D} ) \lesssim 4$. Combination of Eqs.~(\ref{eq:DeltaSC})
and (\ref{eqn:4x4renorm}) yields the practically relevant \textit{s}-wave
pair-potential self-consistency condition
\begin{equation}\label{eq:practSC}
0 = \frac{1}{\Omega}\sum_\kk \left\{ \frac{1}{\Delta}\sum_\eta \left[
u^\uparrow_{\kk - ,\eta} \big(v^\downarrow_{\kk - ,\eta} \big)^* +
u^\downarrow_{\kk + ,\eta} \big( v^\uparrow_{\kk + ,\eta} \big)^* \right]
- \frac{2}{ 2\epsilon_\kk + E_{\mathrm b}} \right\} \quad .
\end{equation}

The densities $n_\sigma$ of spin-$\sigma$ particles are also implicit
functions of system parameters via the expressions
\begin{subequations}
\begin{align}\label{eq:dens}
n_\sigma =& \, \frac{1}{\Omega} \sum_\kk n_{\kk\sigma} \quad , \\
n_{\kk\sigma} =& \, \frac{1}{2} \sum_\eta \left( \big| u^\sigma_{\kk
- ,\eta} \big|^2 + \big| v^\sigma_{\kk + ,\eta} \big|^2 \right) \quad .
\label{eq:densDist}\end{align}
\end{subequations}
For a uniform Fermi gas with fixed total particle number density $n\equiv
\sum_\sigma n_\sigma$, we thus have a second self-consistency
condition given by
\begin{equation}\label{eq:numbSelf}
1 = \frac{1}{\Omega}\sum_\kk \frac{1}{n}  \sum_\sigma n_{\kk\sigma}
\quad .
\end{equation}

Explicit analytical expressions for the momentum-space density
distributions $n_{\kk\sigma}$ defined in (\ref{eq:densDist}) and the
quantity
\begin{equation}\label{eq:Upsilon}
\Upsilon_\kk = \frac{1}{\Delta}\sum_\eta \left[ u^\uparrow_{\kk - ,\eta}
\big(v^\downarrow_{\kk - ,\eta} \big)^* + u^\downarrow_{\kk + ,\eta}
\big( v^\uparrow_{\kk + ,\eta} \big)^* \right]
\end{equation}
entering the r.h.s.\ of Eq.~(\ref{eq:practSC}) have been derived within
the exact $4\times 4$ BdG theory~\cite{Zhou2011,He2012a}. We now
discuss in some detail the corresponding results provided by the
approximate $2\times 2$-projected approach developed here.

\subsection{Momentum-space density distributions and chemical
potential}

\begin{figure}[t]
\includegraphics[width=0.326\textwidth]{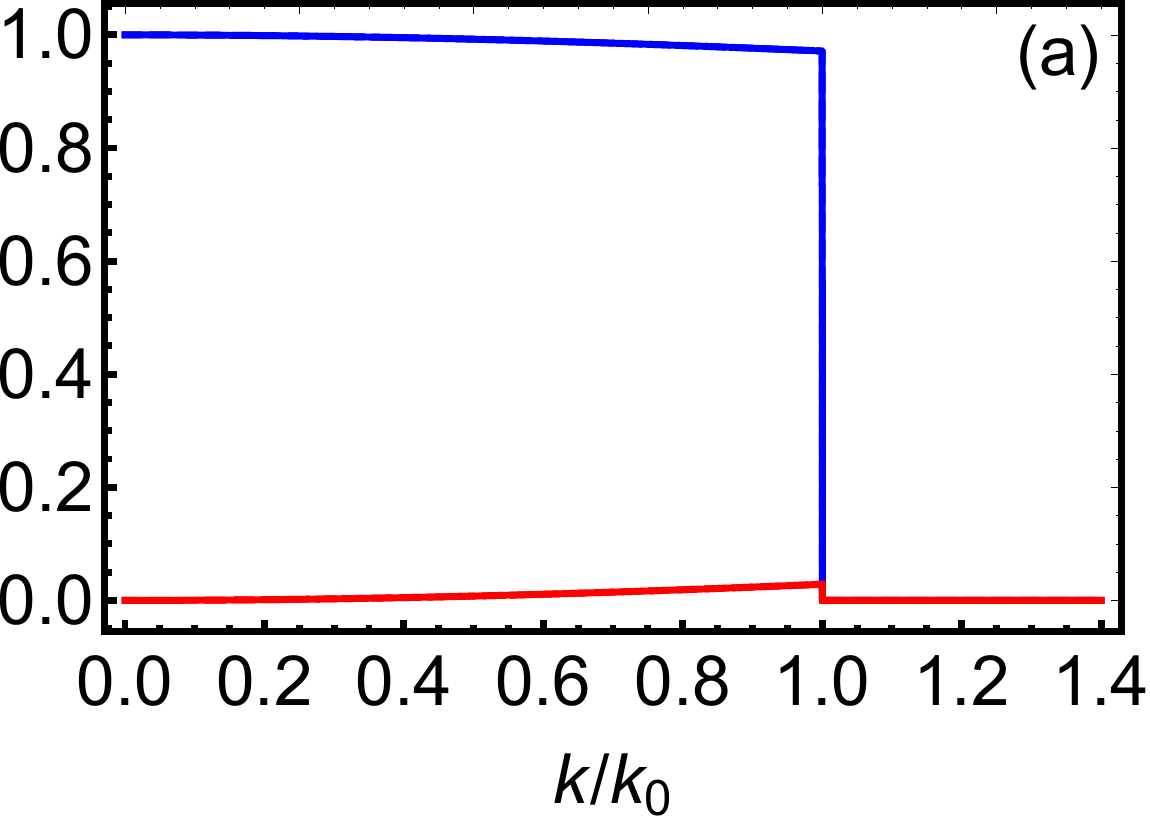}\hfill
\includegraphics[width=0.326\textwidth]{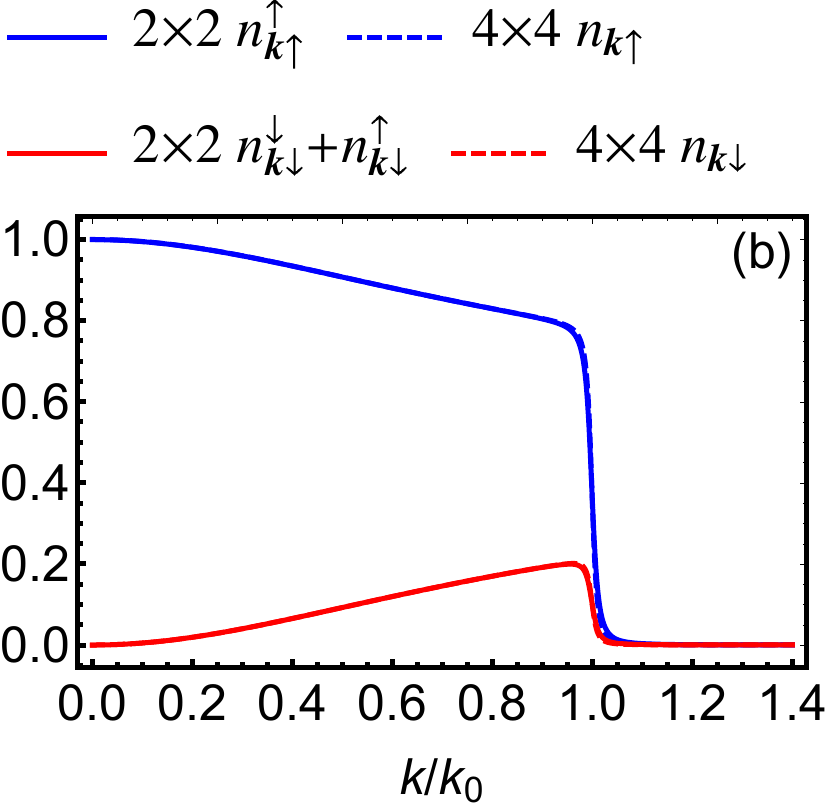}\hfill
\includegraphics[width=0.32\textwidth]{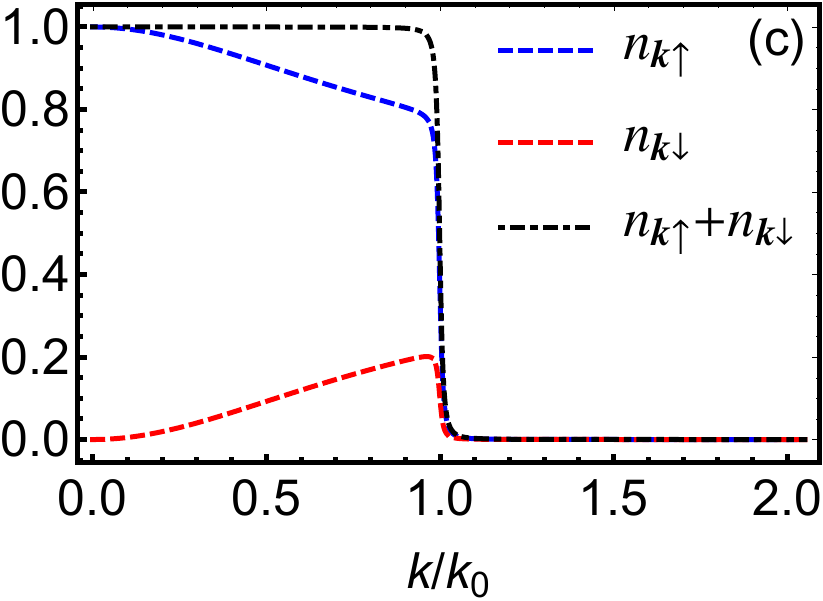}\\[0.2cm]
\includegraphics[width=0.326\textwidth]{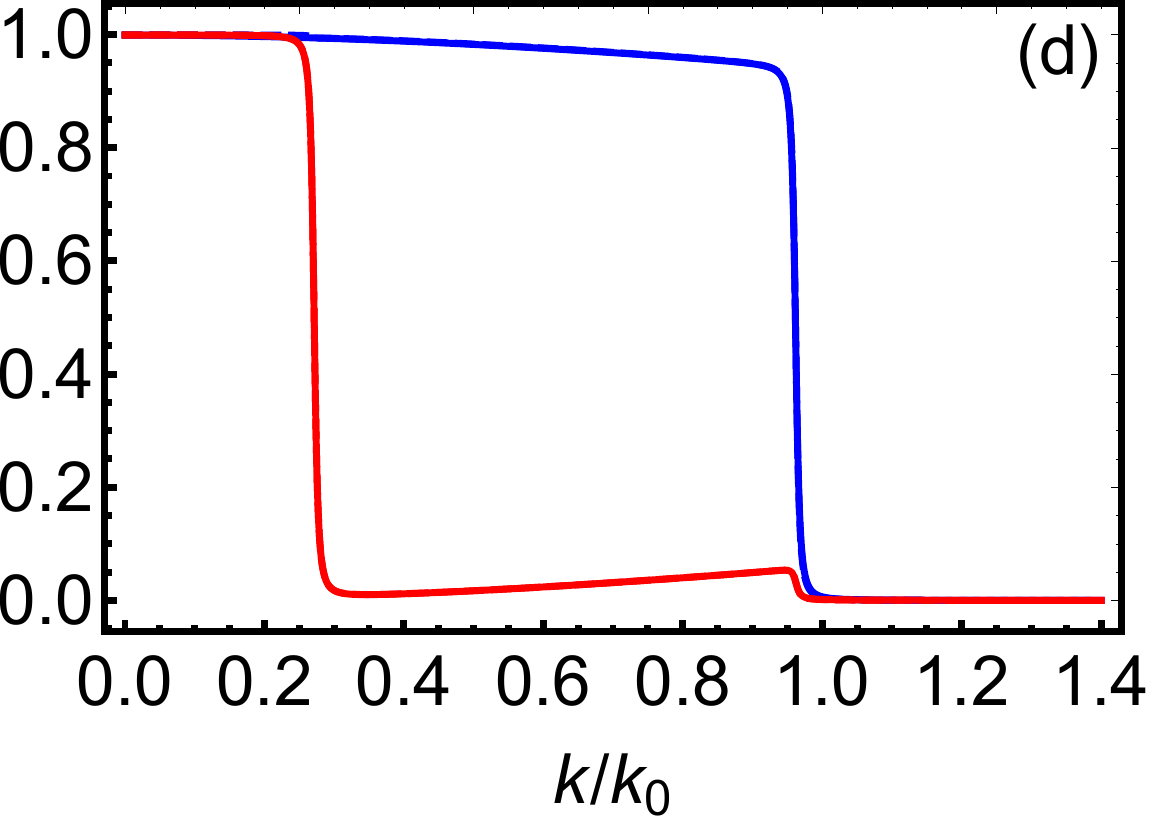}\hfill
\includegraphics[width=0.326\textwidth]{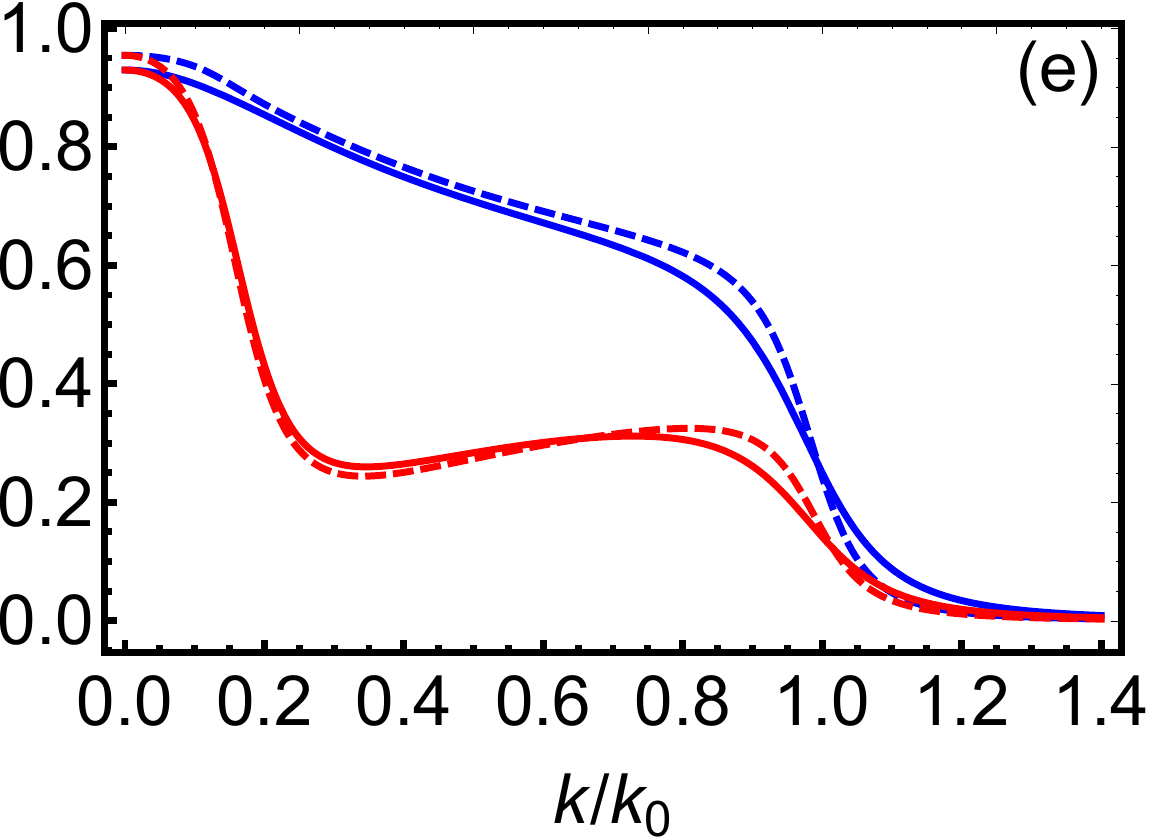}\hfill
\includegraphics[width=0.319\textwidth]{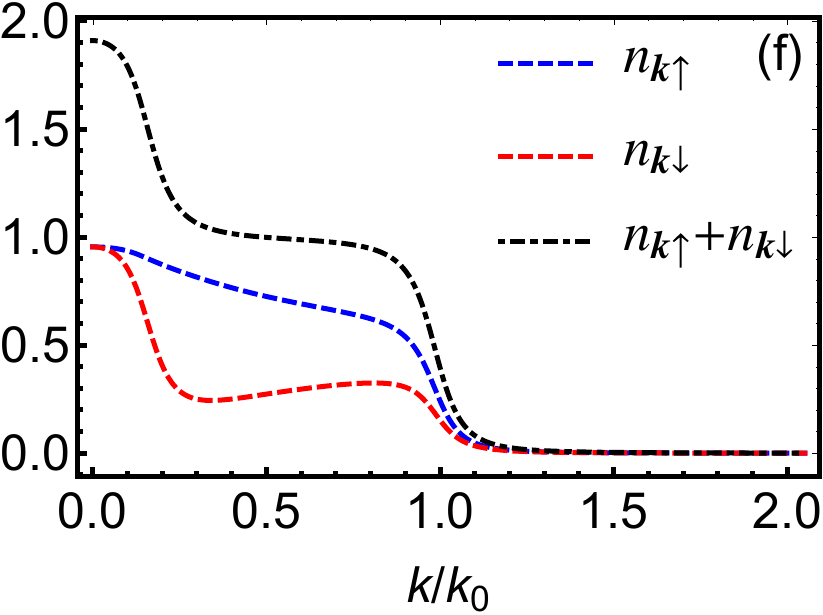}
\caption{\label{fig:densComp}%
Spin-$\sigma$ particle-density distributions $n_{\kk\sigma}$ in
topological superfluids [panels (a-c)] and non-topological superfluids
[panels (d-f)]. Curves labeled $2\times 2$ ($4\times 4$) are obtained
within our $2\times 2$ projected theory, omitting the contribution
$n_{\kk\uparrow}^\downarrow$ to $n_{\kk\uparrow}$ (the exact $4\times
4$ approach). The energy and momentum scales $E_0$ and $k_0$ are
related via $E_0 \equiv \hbar^2 k_0^2/(2 m)$. Panels~(a) and (d) show
situations with excellent agreement between the two approaches. Values
for relevant parameters are $2m\lambda/\hbar^2 k_0 = 0.21$ [in both
panels], $h/E_0 = 0.60$ [in (a)] $0.40$ [in(d)], $\mu/E_0 = 0.36$ [in (a)]
$0.48$ [in (d)], and $|\Delta|/E_0 = 6.3\times 10^{-5}$ [in (a)] $0.024$ [in
(d)], which correspond to self-consistent results obtained for
$E_\mathrm{b}/E_\mathrm{F}=0.50$ when setting $k_0 = \sqrt{2}\,
k_\mathrm{F}$. Deviations between $2\times 2$ and $4\times 4$ results
occur for larger spin-orbit-coupling strength and are more pronounced in
the non-topological regime, as illustrated in panels (b) and (e). Here $2
m \lambda/(\hbar^2 k_0) = 0.71$ [in both (b) and (e)], $h/E_0 = 0.50$ [in
(b)] $0.20$ [in (e)], $\mu/E_0 = 0.13$ [in (b)] $0.24$ [in (e)], and $|\Delta|
/E_0 = 0.019$ [in (b)] $0.11$ [in (e)]. With $k_0=\sqrt{2}\, k_\mathrm{F}$,
these are the parameters obtained self-consistently for $E_\mathrm{b}/
E_\mathrm{F} = 0.050$. Panel (c) [(f)] again displays the exactly
calculated density distributions $n_{\kk\sigma}$ from panel (b) [(e)], with
their sum also shown as the dot-dashed curve.}
\end{figure}

Using results for the spinor amplitudes given in Eq.~(\ref{eq:projSpinor}),
we obtain the momentum-space distribution $n_{\kk\sigma}$ of the
spin-$\sigma$ particle density as a sum of contributions from the two
projected $2\times 2$ sectors, $n_{\kk\sigma} = n_{\kk\sigma}^\sigma
+ n_{\kk\sigma}^{\bar\sigma}$, where
\begin{subequations}
\begin{eqnarray}\label{eq:diagSpinDen}
n_{\kk\sigma}^\sigma &=& \frac{N_{\kk+}^\sigma + N_{\kk-}^\sigma}{4}
\left( 1 - \frac{\xi_{\kk\sigma}}{E_{\kk+}^\sigma} \right) \quad , \\[0.2cm]
n_{\kk\sigma}^{\bar\sigma} &=& \frac{N_{\kk+}^{\bar\sigma} + N_{\kk
-}^{\bar\sigma}}{4} \,\, \frac{|\Delta|^2 + |\lambda_\kk|^2 + (|\Delta|^2 -
|\lambda_\kk|^2) (\xi_{\kk{\bar\sigma}}/E_{\kk+}^{\bar\sigma}) \,
\substack{-\\(+)}\, 2 |\lambda_\kk|^2 | \Delta|^2 / ( h_\kk E_{\kk+}^{\bar
\sigma})}{\left( E_{\kk+}^{\bar\sigma} + \epsilon_{\kk{\sigma}} - \mu
\right)^2} \, , \nonumber \\ \label{eq:upSpinFromDown}
\end{eqnarray}
\end{subequations}
and the upper (lower) sign of the last term in the numerator of
Eq.~\eqref{eq:upSpinFromDown} applies to $\sigma = \,\uparrow$
($\downarrow$). Interestingly, $n_{\kk\uparrow}^{\downarrow}$ turns out
to be negligible except for an unphysical divergence at the point where
the expression in the denominator of Eq.~(\ref{eq:upSpinFromDown}) for
$\sigma = \,\,\uparrow$ vanishes. This artefact of our approximations is
remedied by neglecting $n_{\kk\uparrow}^{\downarrow}$ (i.e., setting
$n_{\kk\uparrow} \equiv n_{\kk\uparrow}^{\uparrow}$ within the $2\times
2$-projected theory) from now on. In contrast, $n_{\kk
\downarrow}^{\uparrow}$ is well-behaved [as the denominator of
Eq.~(\ref{eq:upSpinFromDown}) for $\sigma = \,\,\downarrow$ is always
finite] and contributes importantly to $n_{\kk\downarrow}$.
Figure~\ref{fig:densComp} shows a comparison between the density
distributions $n_{\kk\sigma}$ thus obtained within the $2\times
2$-projected theory with those calculated within the exact $4\times 4$
formalism\footnote{Curves corresponding to exact results obtained from
$4\times 4$ theory in Fig.~\ref{fig:densComp} can be also usefully
compared with the momentum-space density distributions of a 3D Fermi
gas with 2D Rashba spin-orbit coupling calculated for fixed $k_z=0$.
Pertinent results are shown, e.g., as insets of Figs.~3(c) and 3(d) in
Ref.~\cite{Yi2011}.}. There is excellent agreement for the spin-resolved
density distributions from both approaches as long as spin-orbit coupling
is not too strong. For fixed spin-orbit-coupling strength, deviations are
greater for smaller values of the Zeeman splitting $h$, i.e., these tend to
be more pronounced in the non-topological regime.

Interestingly, the concept of separate spin-$\downarrow$ and
spin-$\uparrow$ Fermi spheres with radius $k_\downarrow$ and
$k_\uparrow$, respectively, turns out to be useful even at significant
levels of spin-orbit coupling, since the exact density distributions satisfy
very accurately the approximate relation
\begin{equation}\label{eq:densApprox}
n_{\kk\uparrow} + n_{\kk\downarrow} \approx \Theta(k_\uparrow - k) +
\Theta(k_\downarrow - k) \quad ,
\end{equation}
as is illustrated in Figs.~\ref{fig:densComp}(c) and \ref{fig:densComp}(f).
Here $\Theta(\cdot)$ denotes the Heaviside step function, $k_\uparrow
> k_\downarrow$ generically, and $k_\downarrow \equiv 0$ in the
topological regime. Motivated by observing the apparent broad validity
of relation (\ref{eq:densApprox}), we insert it into the number-density
equation (\ref{eq:numbSelf}) and straightforwardly derive the result
\begin{equation}\label{eq:quickSC}
2 k_\mathrm{F}^2 = k_\uparrow^2 + k_\downarrow^2
\end{equation}
as an equivalent self-consistency condition. Furthermore,
Eq.~(\ref{eq:asympDisp}) together with the fact that $E_{\kk+,<} \approx
0$ for $|\kk|=k_\uparrow$ (generally valid to leading order in small
$|\Delta|$) implies
\begin{subequations}\label{eq:Mu}
\begin{equation}\label{eq:TopoMu}
\mu \approx \frac{\hbar^2 k_\uparrow^2}{2 m} - \sqrt{h^2 + \lambda^2
k_\uparrow^2} \quad .
\end{equation}
For the case where $k_\downarrow\ne 0$, we can extrapolate
Eq.~(\ref{eq:asympDisp}) to the point $E_{\kk+,>} \approx 0$ when
$|\kk| = k_\downarrow$ and find
\begin{equation} \label{eq:TrivMu}
\mu \approx \frac{\hbar^2 k_\downarrow^2}{2 m} + \sqrt{h^2 +
\lambda^2 k_\downarrow^2} \qquad (k_\downarrow > 0) \quad .
\end{equation}
\end{subequations}
In the topological regime (realized for $h > h_\mathrm{c}$, where
$h_\mathrm{c}$ denotes the value for the Zeeman energy at the
transition), $k_\downarrow = 0$ so that Eq.~(\ref{eq:quickSC}) implies
$k_\uparrow = \sqrt{2}\, k_\mathrm{F}$ and (\ref{eq:TopoMu}) yields the
approximate relation
\begin{equation}\label{eq:muInTop}
\frac{\mu}{E_\mathrm{F}} \approx 2 - \sqrt{\left(
\frac{h}{E_\mathrm{F}}\right)^2 + 2 \left( \frac{2 m \lambda}{\hbar^2
k_\mathrm{F}}\right)^2} \qquad (h > h_\mathrm{c})
\end{equation}
between chemical potential and particle density. In the non-topological
regime (realized for $h < h_\mathrm{c}$), $k_\downarrow\ne 0$ and we
need to simultaneously solve Eqs.~(\ref{eq:TopoMu}), (\ref{eq:TrivMu})
and (\ref{eq:quickSC}). Adding (\ref{eq:TrivMu}) to (\ref{eq:TopoMu}),
using (\ref{eq:quickSC}), and expanding to first sub-leading order in
large $h$, we find
\begin{subequations}
\begin{equation}\label{eq:TrivMuStep}
\frac{\mu}{E_\mathrm{F}} \approx 1 - \frac{m \lambda^2}{2 \hbar^2 h}\,
\frac{k_\uparrow^2 - k_\downarrow^2}{k_\mathrm{F}^2} \qquad (h <
h_\mathrm{c}) \quad .
\end{equation}
Furthermore, subtracting (\ref{eq:TrivMu}) from (\ref{eq:TopoMu}) and
expanding again to first sub-leading order in large $h$ yields
\begin{equation}\label{eq:densDiff}
\frac{k_\uparrow^2 - k_\downarrow^2}{k_\mathrm{F}^2} \approx \frac{2
h}{E_\mathrm{F}} + \frac{2 m \lambda^2}{\hbar^2 h} \qquad (h <
h_\mathrm{c}) \quad ,
\end{equation}
\end{subequations}
where we have again also used Eq.~(\ref{eq:quickSC}). Combining the
results from Eqs.~(\ref{eq:TrivMuStep}) and (\ref{eq:densDiff}), we
find the relation
\begin{equation}\label{eq:muInTriv}
\frac{\mu}{E_\mathrm{F}} \approx 1 - \frac{1}{2} \left( \frac{2 m
\lambda}{\hbar^2 k_\mathrm{F}} \right)^2 \qquad (h < h_\mathrm{c})
\end{equation}
between chemical potential and particle number that is valid in the
non-topological regime, to leading order in large $h$.

\begin{figure}[t]
\centerline{%
\includegraphics[height=4.03cm]{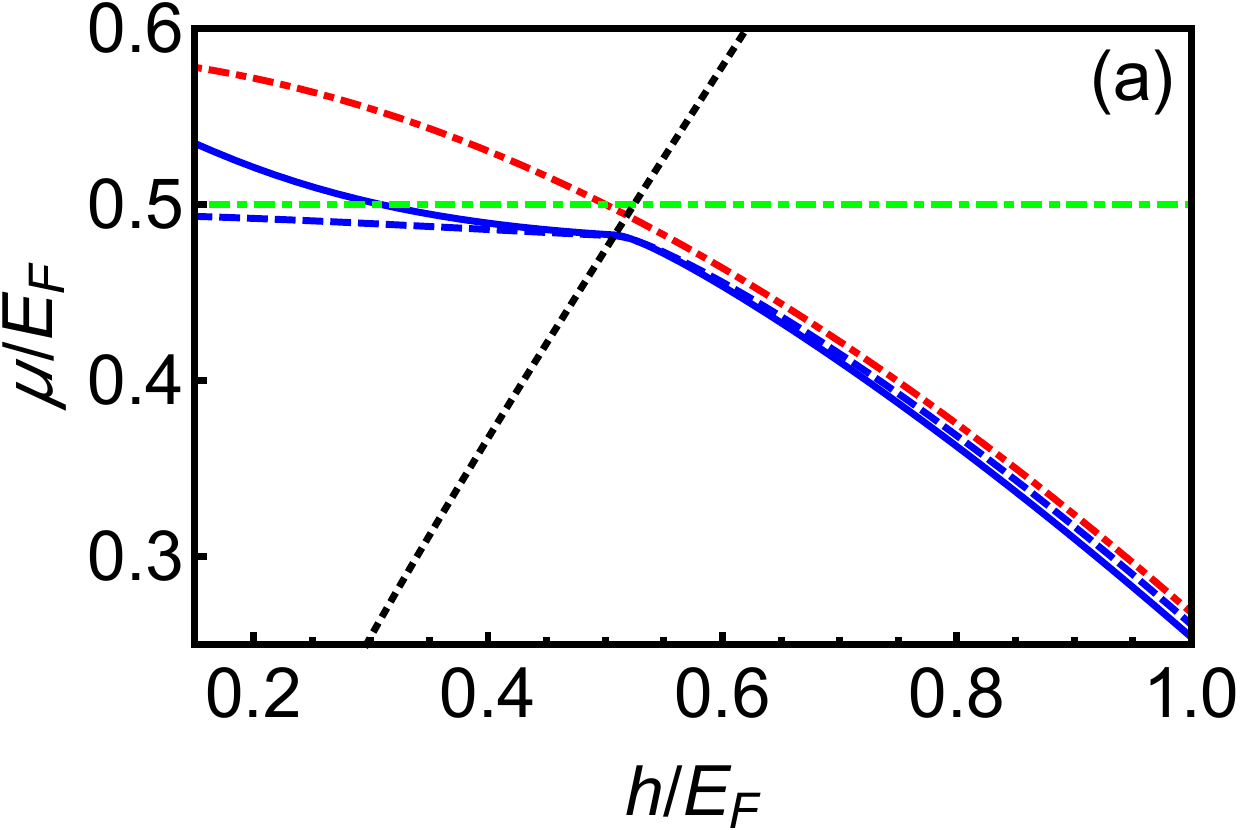}\hfill
\includegraphics[height=4cm]{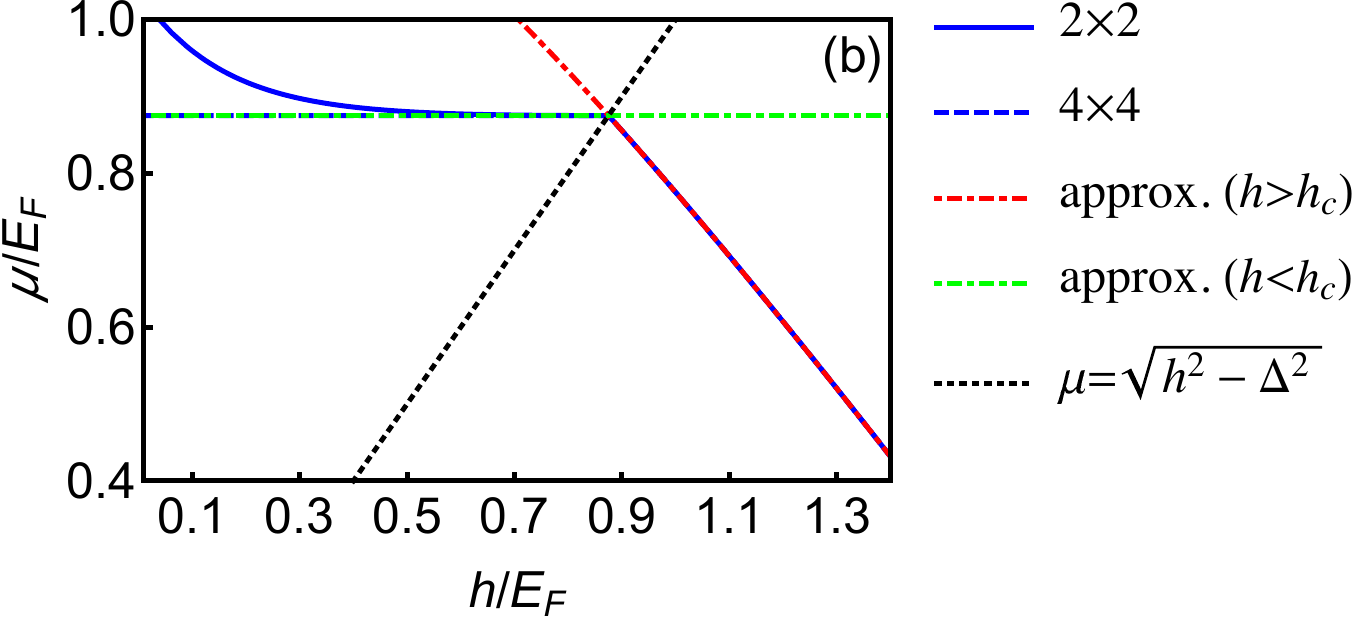}
}%
\caption{\label{fig:muCompare}%
Variation of the chemical potential $\mu$ with Zeeman energy $h$ for
fixed total density $n = m E_\mathrm{F}/(\pi\hbar^2)$. Curves labeled
$2\times 2$ ($4\times 4$) are obtained using our $2\times 2$-projected
theory, omitting the contribution $n_{\kk\uparrow}^\downarrow$ to
$n_{\kk\uparrow}$ (the exact $4\times 4$ approach). Panel (a) [(b)]
shows results calculated for $2m\lambda/(\hbar^2 k_\mathrm{F}) =
1.00$ [$0.500$]. For convenience, we fixed $|\Delta|/E_\mathrm{F} =
0.159$ [$3.48\times 10^{-4}$] in the calculation, which is the
self-consistent value at the critical Zeeman energy $h_\mathrm{c}/
E_\mathrm{F} = 0.507$ [$0.875$] for $E_\mathrm{b}/E_\mathrm{F} =
0.0462$, i.e., $\ln ( k_\mathrm{F} a_\mathrm{2D} ) = 2.00$. The
approximate analytical formulae from Eqs.~(\ref{eq:muInTop}) and
(\ref{eq:muInTriv}) are plotted as the dot-dashed curves, and the dotted
curve indicates the condition for the transition between non-topological
and topological superfluid phases.}
\end{figure}

Figure~\ref{fig:muCompare} shows a detailed comparison between the
self-consistent chemical potential obtained from the exact $4\times 4$
approach, from the effective $2\times 2$-projected theory developed
here, and the approximate analytical expressions (\ref{eq:muInTop}) and
(\ref{eq:muInTriv}). The situation depicted in panel~(a) is the same as in
Fig.~1(b) of Ref.~\cite{He2012a}. Capitalizing on the weak $|\Delta|$
dependence, we used fixed values for the \textit{s}-wave gap in our
calculation, corresponding to the self-consistent results at $h =
h_\mathrm{c}$ for $E_\mathrm{b}/E_\mathrm{F} = 0.0462$. The curve
for the $2\times 2$-projected theory is seen to agree very well with the
exact result for large-enough $h/E_\mathrm{F}$, which includes not only
the topological regime but also part of the non-topological regime near
the transition. Deviations between the $2\times 2$-projected theory and
the exact $4\times 4$ results become significant in the limit of small $h$,
where the Feshbach-projection approach is indeed expected to fail. As
illustrated by the situation shown in panel~(b), the agreement between
the effective-$2\times 2$ and exact-$4\times 4$ results becomes
excellent for smaller values of $\lambda$, reaching also deeper into the
non-topological regime. The observation that the exact $4\times
4$-theory results for $\mu/E_\mathrm{F}$ and the approximate
analytical expressions given in Eqs.~(\ref{eq:muInTop}) and
(\ref{eq:muInTriv}) are practically indistinguishable at small-enough
magnitude of spin-orbit coupling suggests the possibility to utilize these
analytical formulae, for better efficiency and greater insight, as input into
the self-consistent calculation of the \textit{s}-wave pair potential. The
analytical expression (\ref{eq:muInTriv}) could also be useful to more
accurately represent the chemical potential in the low-$h$ limit where
the $2\times 2$-projected results deviate significantly from those
obtained from the exact $4\times 4$ approach.

\subsection{Self-consistency of \textit{s}-wave pair potential:
Spin-$\uparrow$-projected theory}

The approximate description based on the $2\times 2$-subspace
projections gave Eq.~(\ref{eq:projSpinor}) for the Nambu-spinor
amplitudes. Inserting these expressions into (\ref{eq:Upsilon}) yields
$\Upsilon_\kk = \Upsilon_\kk^\uparrow + \Upsilon_\kk^\downarrow$, with
\begin{equation}\label{eqn:gapnum}
\Upsilon_\kk^{\uparrow(\downarrow)} = \frac{N_{\kk+}^{\uparrow
(\downarrow)} \,\substack{+\\(-)}\,\, N_{\kk-}^{\uparrow(\downarrow)}}{2
E_{\kk+}^{\uparrow(\downarrow)}}\,\, \frac{\epsilon_{\kk{\uparrow
(\downarrow)}} -\mu\,\substack{-\\(+)}\, E_{\kk+}^{\uparrow(\downarrow)}
\substack{+\\(-)}\, (|\Delta|^2+|\lambda_\kk|^2)/(2 h_\kk)}{\epsilon_{\kk
\downarrow(\uparrow)} - \mu \, \substack{-\\(+)}\, E_{\kk+}^{\uparrow
(\downarrow)}} \quad .
\end{equation}
Using our approximation (\ref{eq:normUp}) for the normalization factors
and, for consistency, replacing $\epsilon_{\kk\downarrow} -\mu - E_{\kk
+}^\uparrow\approx 2 h_\kk$ in the denominator of
$\Upsilon_\kk^\uparrow$, we obtain
\begin{subequations}
\begin{eqnarray}\label{UpsUpApp}
\Upsilon_\kk^\uparrow &\approx& \frac{1}{E_{\kk+}^\uparrow}\, \frac{2
h_\kk\left( \epsilon_{\kk\uparrow}-\mu - E_{\kk+}^\uparrow \right) +
|\Delta|^2 + |\lambda_\kk|^2}{4 h_\kk^2 + |\Delta|^2 + |\lambda_\kk|^2}
\quad , \\[0.2cm] \label{eq:approxYpsDwn}
\Upsilon_\kk^\downarrow &\approx& 0 \quad .
\end{eqnarray}
\end{subequations}
Hence, we find that the self-consistency condition for the \textit{s}-wave
pair potential can be formulated entirely in terms of quantities relating to
the projected spin-$\uparrow$ degrees of freedom.

\begin{figure}[t]
\includegraphics[width=0.33\textwidth]{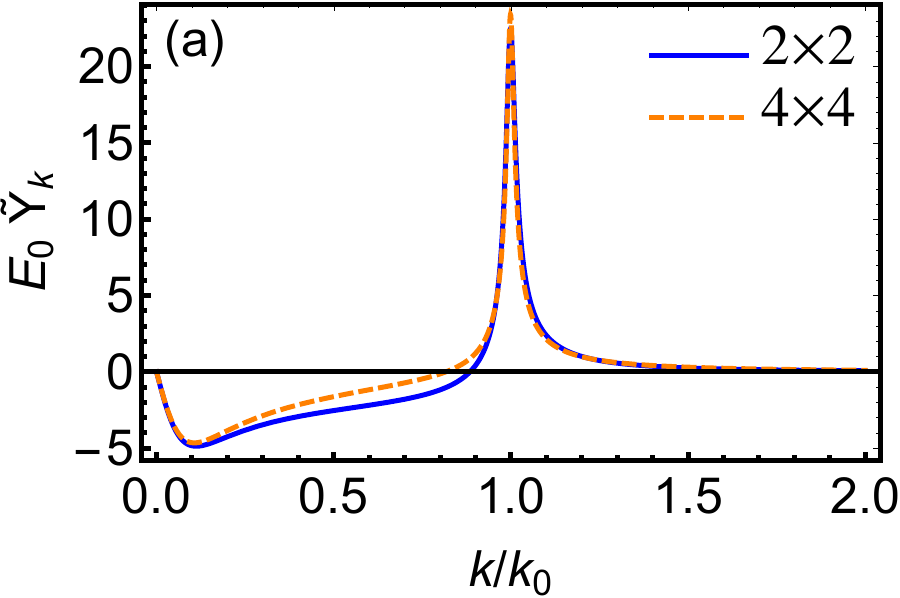}\hfill
\includegraphics[width=0.33\textwidth]{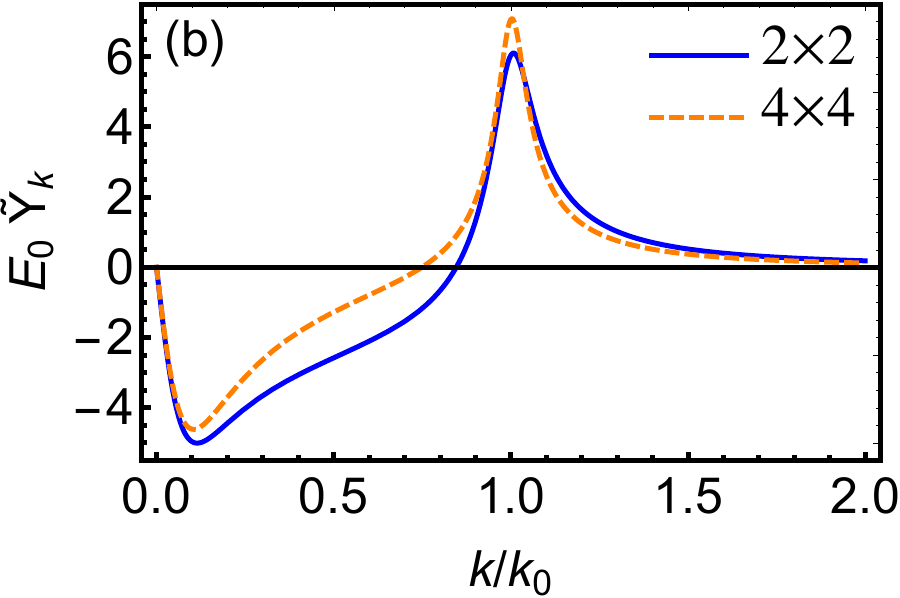}\hfill
\includegraphics[width=0.33\textwidth]{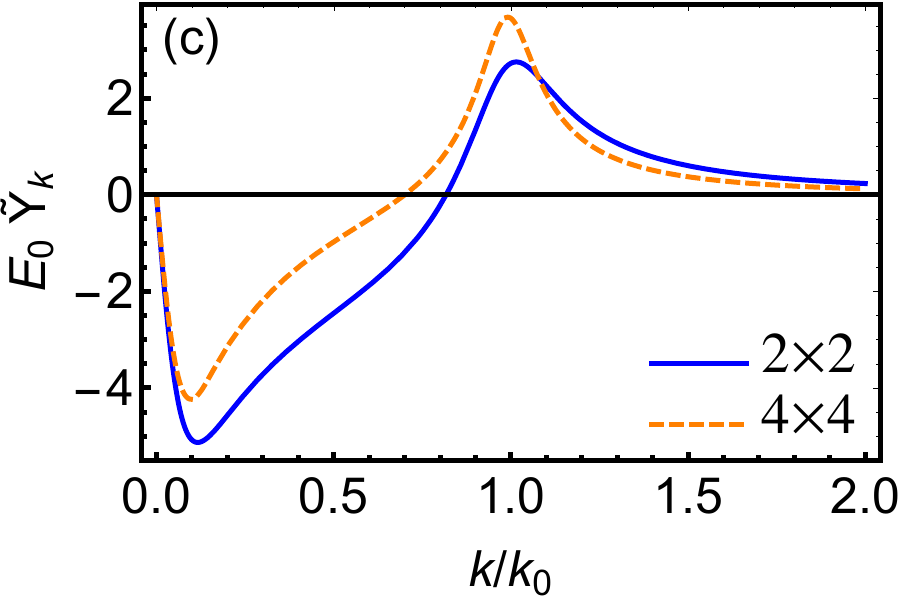}
\caption{\label{fig:gapIntComp}%
Line shape of the summand $\tilde\Upsilon_\kk = \Upsilon_\kk - 2/(2
\epsilon_\kk + E_\mathrm{b})$ in the self-consistency condition
(\ref{eq:practSC}) for the \textit{s}-wave pair potential. Curves labeled
$2\times 2$ ($4\times 4$) are calculated from the $2\times 2$-projected
theory using $\Upsilon_\kk \equiv \Upsilon_\kk^\uparrow$ with
(\ref{UpsUpApp}) (from the exact $4\times 4$-theory expression for
$\tilde\Upsilon_\kk$). The parameters $E_\mathrm{b}/E_0 = 0.023$ and
$2 m\lambda/(\hbar^2 k_0) = 0.71$ are fixed in all panels, whereas $h/
E_0 = 0.50$ [$0.30$, $0.20$], $\mu/E_0 = 0.13$ [$0.23$, $0.24$], and
$|\Delta|/E_0 = 0.017$ [$0.060$, $0.11$] for panel (a) [(b), (c)].  With
$k_0 = \sqrt{2}\, k_\mathrm{F}$, these values coincide with those used
for/obtained by numerical calculations whose results are shown in
Fig.~1 of Ref.~\cite{He2012a}. The energy and momentum scales
$E_0$ and $k_0$ are related via $E_0\equiv\hbar^2 k_0^2/(2 m)$.}
\end{figure}

Figure~\ref{fig:gapIntComp} shows the $k$ dependence of the quantity
$\tilde\Upsilon_\kk = \Upsilon_\kk - 2/(2 \epsilon_\kk + E_\mathrm{b})$
that is the summand in the self-consistency equation (\ref{eq:practSC})
for the \textit{s}-wave pair potential. Parameters are chosen to coincide
with those from a recent numerical study~\cite{He2012a}. The system is
deep in the topological-superfluid regime for panel (a), still topological
but close to the transition in panel (b), and a non-topological superfluid
close to the transition in panel (c). Perhaps not surprisingly, the
agreement between the projected $2\times 2$ theory and the exact
$4\times 4$ formalism is best deep in the topological-superfluid regime,
as the fidelity of the projected theory should improve for increasing $h$.
Generally, the small-$k$ and large-$k$ behaviors of $\tilde\Upsilon_\kk$
are captured almost perfectly within the projected $2\times 2$ theory,
with deviations at smaller $h$ occurring chiefly at intermediate values of
$k$. However, as the self-consistency condition (\ref{eq:practSC})
involves a sum over all $k$, the overall effect of such deviations cannot
be easily ascertained without explicitly finding the self-consistent
\textit{s}-wave pair potentials within both the $4\times 4$ and $2\times
2$ approaches. Such a detailed comparison is one of the foci of the
next Section.

\section{Superfluidity with uniform \textit{s}-wave pair potential:
Effective two-band description \textit{versus\/} exact four-band theory}
\label{sec:TheoComp}

The complete description of superfluidity for a uniform system in the
experimentally relevant situation with fixed total particle density requires
the simultaneous solution of the self-consistency conditions
(\ref{eq:practSC}) and (\ref{eq:numbSelf}). This is generally achieved by
standard iterative procedures that are based on a fully explicit
knowledge of the exact four-band spectrum $E_{\kk\alpha,\eta}$ and the
associated eigenspinors. Although such a procedure has the advantage
of yielding exact results, its complicated formal structure obscures
possibilities for gaining a deeper intuitive understanding of relevant
physical effects. In contrast, the formalism developed in
Sec.~\ref{sec:selfCon} offers the attractive alternative to be able to
describe the system entirely in terms of a conceptually simpler theory
based on the spin-$\uparrow$-projected (two-band) spectrum. We now
investigate in greater detail the physical picture provided by the effective
two-band approach, where the self-consistency condition
(\ref{eq:practSC}) is solved using $\Upsilon_\kk \equiv
\Upsilon_\kk^\uparrow$ with (\ref{UpsUpApp}) and approximating the
chemical potential by Eqs.~(\ref{eq:muInTop}) and (\ref{eq:muInTriv}) as
appropriate. 

\subsection{Boundary between topological and non-topological phases}

We start by considering relevant thermodynamic quantities at the
transition between the non-topological and topological superfluid
regimes. This transition occurs at the value $h\equiv h_\mathrm{c}$ of
the Zeeman energy that satisfies the condition
\begin{equation}\label{eq:hCrit}
h_\mathrm{c} = \sqrt{\mu^2 + |\Delta|^2} \quad .
\end{equation}
For a given system with fixed uniform total particle density $n$ and
\textit{s}-wave interaction strength measured in terms of the two-body
bound-state energy $E_\mathrm{b}$, both $\mu$ and $|\Delta|$ are
implicit functions of $h$ and $n$ via the self-consistency conditions and,
thus, their values $\mu_\mathrm{c}\equiv \mu(h_\mathrm{c})$ and
$\Delta_\mathrm{c} \equiv |\Delta(h_\mathrm{c})|$ are also fixed. In
Table~\ref{tab:scales}, we summarize these critical values obtained
using the exact $4\times 4$ theory and compare with those calculated
within the effective $2\times 2$ approach using two different methods.
To obtain $\mu_\mathrm{c}^{2\times 2}$ and $\Delta_\mathrm{c}^{2
\times 2}$, we simultaneously solve the self-consistency conditions for
the number density and pair potential, Eqs.~(\ref{eq:numbSelf}) and
(\ref{eq:practSC}), assuming also $n_{\kk\uparrow}\equiv n_{\kk
\uparrow}^\uparrow$, $n_{\kk\downarrow}\equiv n_{\kk
\downarrow}^\downarrow + n_{\kk\downarrow}^\uparrow$, and
$\Upsilon_\kk\equiv \Upsilon_\kk^\uparrow$ with relevant expressions
given in Eqs.~(\ref{eq:diagSpinDen}), (\ref{eq:upSpinFromDown}), and
(\ref{UpsUpApp}). In contrast, $\tilde\Delta_\mathrm{c}^{2\times 2}$ is
the result of a simpler routine where only the pair-potential
self-consistency condition (\ref{eq:practSC}) is solved, setting
$\Upsilon_\kk \equiv \Upsilon_\kk^\uparrow$ with Eq.~(\ref{UpsUpApp})
and approximating $\mu_\mathrm{c}/E_\mathrm{F}$ by
Eq.~(\ref{eq:muInTriv}).

\begin{table}[t]
\caption{\label{tab:scales}
Chemical potential $\mu(h_\mathrm{c}) \equiv \mu_\mathrm{c}$ and
\textit{s}-wave gap $|\Delta(h_\mathrm{c})| \equiv \Delta_\mathrm{c}$ at
the critical Zeeman energy $h_\mathrm{c}$ where the transition between
topological and non-topological regimes occurs, calculated exactly within
$4\times 4$ theory and compared with results from the effective $2\times
2$ approach ($\mu_\mathrm{c}^{2\times 2}$, $\Delta_\mathrm{c}^{2
\times 2}$). The value $\tilde\Delta_\mathrm{c}^{2\times 2}$ is the critical 
gap obtained from $2\times 2$ theory when $\mu_\mathrm{c}/
E_\mathrm{F}$ is approximated by Eq.~(\ref{eq:muInTriv}). Values for
$\mu_\mathrm{c}$ that agree with Eq.~(\ref{eq:muInTriv}) to within 5\%
are shown in green. Results for $\tilde\Delta_\mathrm{c}^{2\times 2}$
given in magenta agree with $\Delta_\mathrm{c}$ to within 25\%.}
\vspace{-0.4cm}
\renewcommand{\arraystretch}{1.04}
\begin{center}
\begin{tabular*}{\columnwidth}{c|c|rrrrrr} \hline \hline
\multirow{2}{*}{$2 m \lambda/(\hbar^2 k_\mathrm{F})$}
& $\ln(k_\mathrm{F} a_\mathrm{2D})$ & $0.500$ & $1.00$ & $1.50$ &
$2.00$ & $2.50$ & $3.00$ \\
& $E_\mathrm{b}/E_\mathrm{F}$ & $0.928$ & $0.341$ & $0.126$ &
$0.0462$ & $0.0170$ & $0.00625$ \\ \hline
\multirow{6}{*}{1.50}
& $h_\mathrm{c}/E_\mathrm{F}$ & $1.32$ & $0.802$ & $0.524$ &
$0.357$ & $0.253$ & $0.189$ \\
& $\mu_\mathrm{c}/E_\mathrm{F}$ & $-0.660$ & $-0.323$ & $-0.188$
& $-0.137$ & \goo{$-0.120$} & $-0.117$ \\
& $\Delta_\mathrm{c}/E_\mathrm{F}$ & $1.14$ & $0.735$ & $0.490$
& $0.330$ & $0.222$ & $0.148$ \\
& $\mu_\mathrm{c}^{2\times 2}/E_\mathrm{F}$ & $-0.348$ & $-0.207$
& $-0.132$ & $-0.0943$ & $-0.0782$ & $-0.0743$ \\
& $\Delta_\mathrm{c}^{2\times 2}/E_\mathrm{F}$ & $0.775$ & $0.582$
&$0.444$ & $0.344$ & $0.269$ & $0.213$ \\
& $\tilde\Delta_\mathrm{c}^{2\times 2}/E_\mathrm{F}$ & $0.792$ &
\foo{$0.586$} & \foo{$0.445$} & \foo{$0.343$} & \foo{$0.269$} &
$0.212$ \\ \hline
\multirow{6}{*}{1.25}
& $h_\mathrm{c}/E_\mathrm{F}$ & $1.15$ & $0.689$ & $0.457$ &
$0.331$ & $0.266$ & $0.236$ \\
& $\mu_\mathrm{c}/E_\mathrm{F}$ & $-0.291$ & $0.0234$ & $0.144$
& $0.190$ & \goo{$0.208$} & \goo{$0.214$} \\
& $\Delta_\mathrm{c}/E_\mathrm{F}$ & $1.11$ & $0.689$ & $0.434$
& $0.271$ & $0.166$ & $0.0982$ \\
& $\mu_\mathrm{c}^{2\times 2}/E_\mathrm{F}$ & $0.0940$ & $0.173$
& $0.207$ & \goo{$0.221$} & \goo{$0.225$} & \goo{$0.225$} \\
& $\Delta_\mathrm{c}^{2\times 2}/E_\mathrm{F}$ & $0.609$ & $0.433$
&$0.313$ & $0.227$ & $0.166$ & $0.122$ \\
& $\tilde\Delta_\mathrm{c}^{2\times 2}/E_\mathrm{F}$ & $0.590$ &
$0.425$ & $0.310$ & \foo{$0.228$} & \foo{$0.168$} & \foo{$0.124$} \\
\hline
\multirow{6}{*}{1.00}
& $h_\mathrm{c}/E_\mathrm{F}$ & $1.07$ & $0.685$ & $0.546$ &
$0.507$ & $0.501$ & $0.500$ \\
& $\mu_\mathrm{c}/E_\mathrm{F}$ & $0.0291$ & $0.330$ & $0.442$ &
\goo{$0.482$} & \goo{$0.495$} & \goo{$0.499$} \\
& $\Delta_\mathrm{c}/E_\mathrm{F}$ & $1.07$ & $0.600$ & $0.320$ &
$0.159$ & $0.0758$ & $0.0358$ \\
& $\mu_\mathrm{c}^{2\times 2}/E_\mathrm{F}$ & $0.439$ & $0.472$ &
\goo{$0.487$} & \goo{$0.494$} & \goo{$0.498$} & \goo{$0.499$} \\
& $\Delta_\mathrm{c}^{2\times 2}/E_\mathrm{F}$ & $0.352$ & $0.215$
& $0.131$ & $0.0794$ & $0.0481$ & $0.0291$ \\
& $\tilde\Delta_\mathrm{c}^{2\times 2}/E_\mathrm{F}$ & $0.329$ &
$0.206$ & $0.128$ & $0.0782$ & $0.0477$ & \foo{$0.0290$} \\ \hline
\multirow{6}{*}{0.75}
& $h_\mathrm{c}/E_\mathrm{F}$ & 1.04 & $0.747$ & $0.718$ &
$0.718$ & $0.719$ & $0.719$ \\
& $\mu_\mathrm{c}/E_\mathrm{F}$ & 0.317 & $0.630$ & \goo{$0.707$}
& \goo{$0.717$} & \goo{$0.719$} & \goo{$0.719$} \\
& $\Delta_\mathrm{c}/E_\mathrm{F}$ & 0.990 & $0.401$ & $0.127$ &
$0.0401$ & $0.0128$ & $0.00409$ \\
& $\mu_\mathrm{c}^{2\times 2}/E_\mathrm{F}$ & \goo{$0.713$} &
\goo{$0.718$} & \goo{$0.718$} & \goo{$0.719$} & \goo{$0.719$} &
\goo{$0.719$} \\
& $\Delta_\mathrm{c}^{2\times 2}/E_\mathrm{F}$ & $0.0702$ &
$0.0287$ & $0.0118$ &$0.00484$ & $0.00199$ & $0.000818$ \\
& $\tilde\Delta_\mathrm{c}^{2\times 2}/E_\mathrm{F}$ & $0.0687$ &
$0.0285$ & $0.0118$ & $0.00484$ & $0.00199$ & $0.000818$ \\ \hline
\hline
\end{tabular*}\end{center}
\end{table}

Inspection of Table~\ref{tab:scales} shows that the values obtained for
$\Delta_\mathrm{c}^{2\times 2}$ and $\tilde\Delta_\mathrm{c}^{2\times
2}$ are generally very close, even in the regime where the
approximation (\ref{eq:muInTriv}) for $\mu_\mathrm{c}/E_\mathrm{F}$ is
not accurate. [For easy reference, values for $\mu_\mathrm{c}/
E_\mathrm{F}$ and $\mu_\mathrm{c}^{2\times 2}/E_\mathrm{F}$ that
agree to within 5\% with the analytical approximation
Eq.~(\ref{eq:muInTriv}) are indicated in green.] Thus, at least to
determine critical values within the effective $2\times 2$-projected
theory, using the simpler routine yielding $\tilde\Delta_\mathrm{c}^{2
\times 2}$ is a viable approach. Interestingly, the agreement between
values for $\Delta_\mathrm{c}$ and $\tilde\Delta_\mathrm{c}^{2\times
2}$ turns out to be generally better for larger $\lambda$. (Values for
$\tilde\Delta_\mathrm{c}^{2\times 2}$ indicated in magenta are close to
within 25\% to the exact $4\times 4$ results.) More specifically, even
though the assumption (\ref{eq:muInTriv}) made for $\mu$ when
determining $\tilde\Delta_\mathrm{c}$ is more broadly valid across the
range of accessible $E_\mathrm{b}$ at smaller $\lambda$, the projected
$2\times 2$-theory's self-consistency equations appear to fail for small
$|\Delta|$. As a rule of thumb, the condition $ 2 m \lambda/(\hbar^2
k_\mathrm{F})\gtrsim 1$ is needed for $2\times 2$-theory results to be in
reasonable agreement with the exact $4\times 4$ values
$\Delta_\mathrm{c}$. Surprisingly, at larger $\lambda$, the rather good
agreement between $\Delta_\mathrm{c}$ and $\tilde
\Delta_\mathrm{c}^{2\times 2}$ extends even to situations where
$\mu_\mathrm{c}$ differs significantly from the approximation
(\ref{eq:muInTriv}).

\begin{figure}[t]
\includegraphics[width=0.325\textwidth]{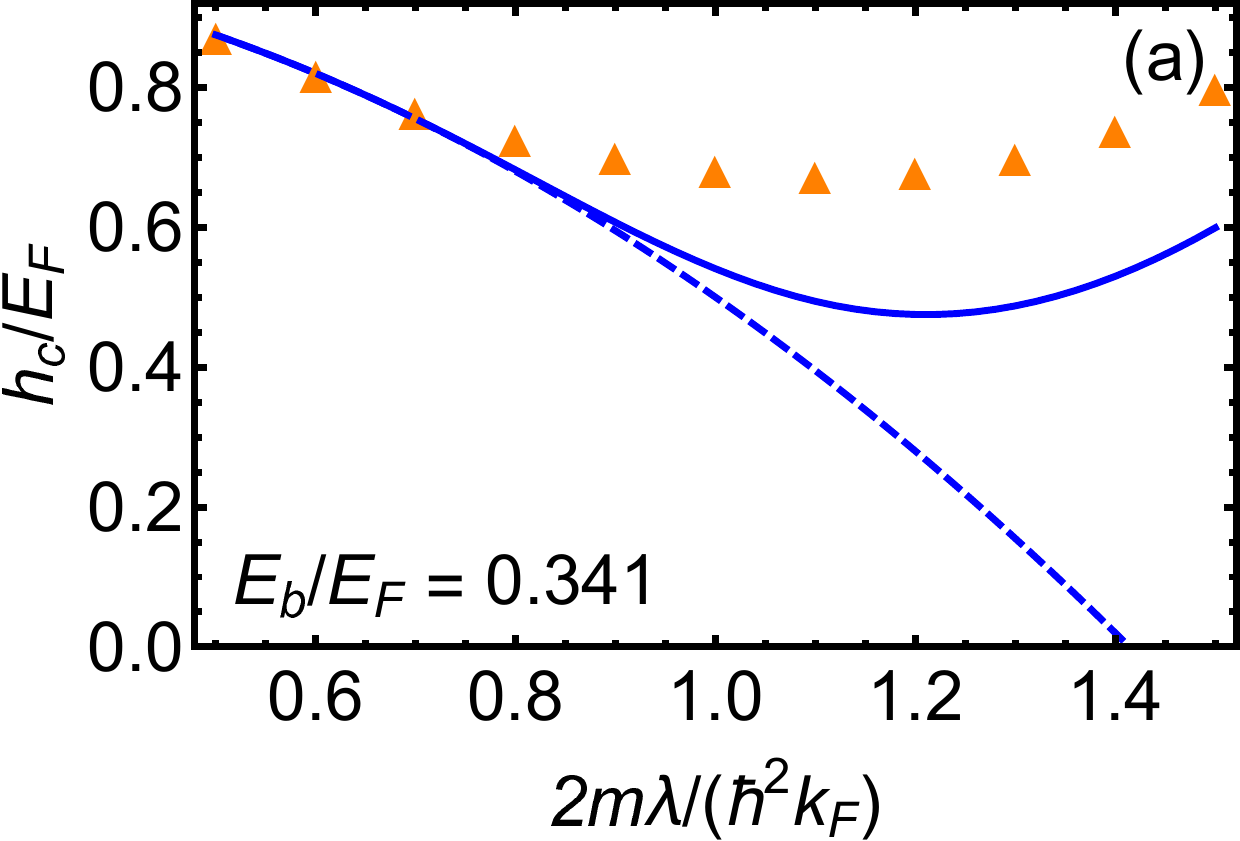}\hfill
\includegraphics[width=0.328\textwidth]{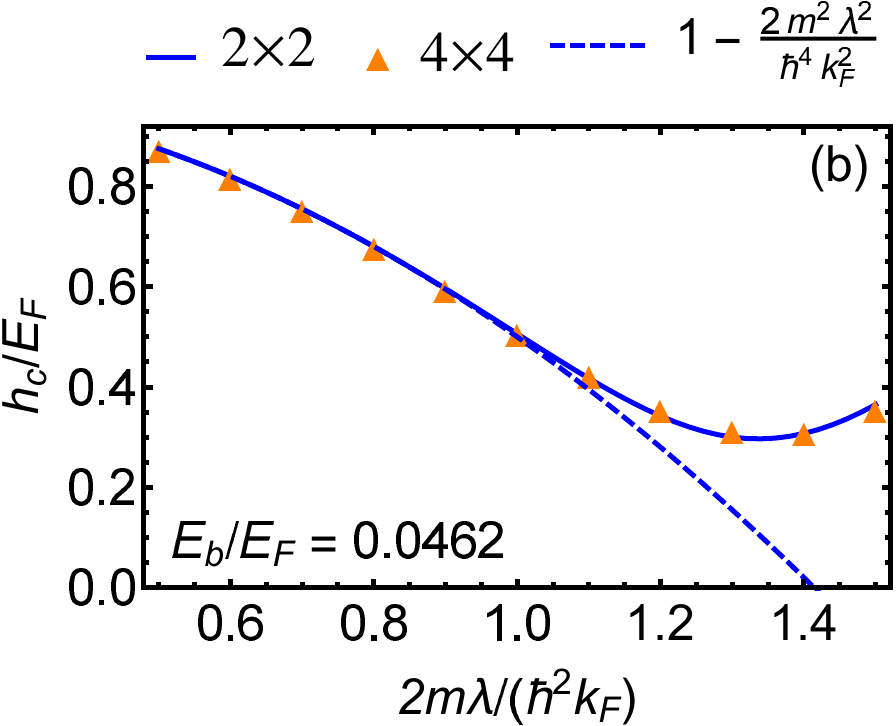}\hfill
\includegraphics[width=0.325\textwidth]{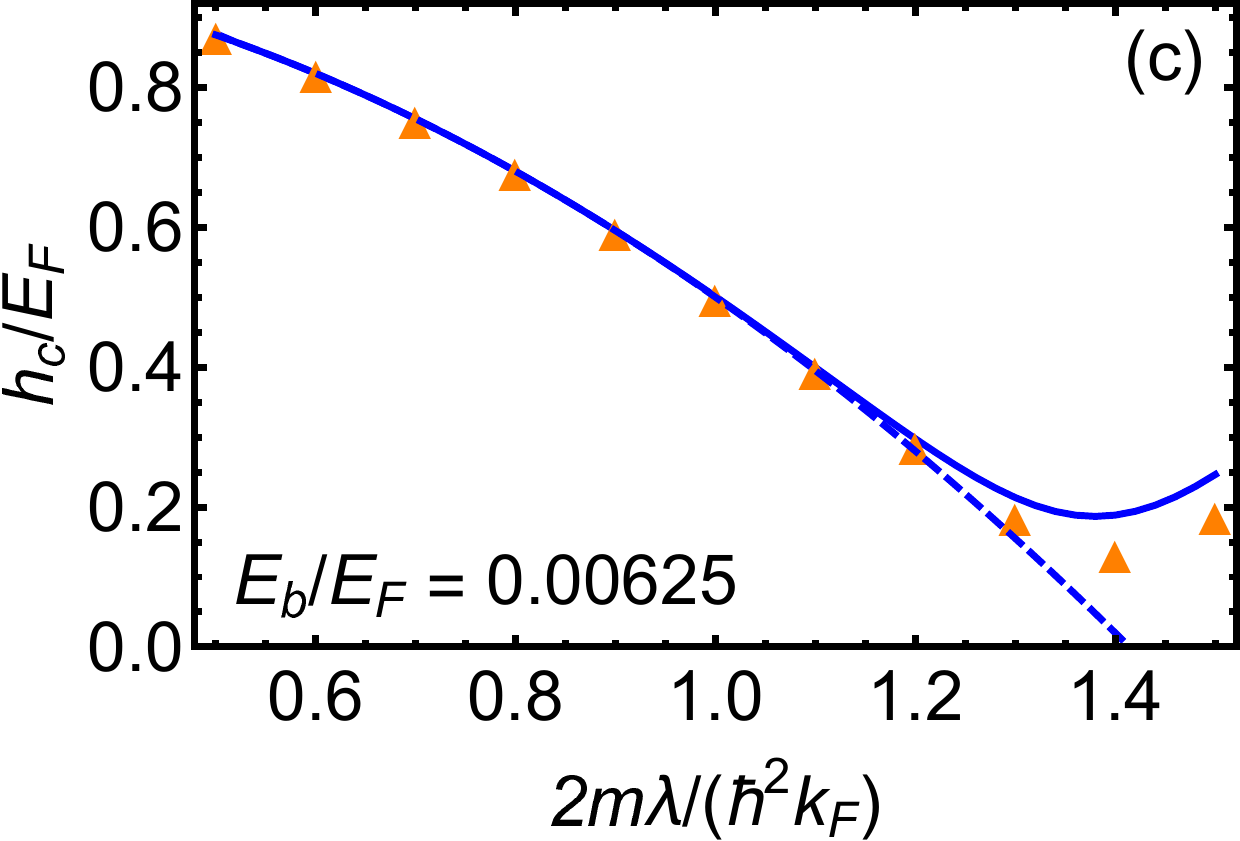}
\caption{\label{fig:phase_b}%
Phase boundary between topological and non-topological superfluid
states that occur for $h > h_\mathrm{c}$ and $h < h_\mathrm{c}$,
respectively. Results labeled $2\times 2$ ($4\times 4$) were calculated
by finding the Zeeman energy satisfying (\ref{eq:hCrit}) from solution of
the self-consistency equation (\ref{eq:practSC}) for the \textit{s}-wave
pair potential using $\Upsilon_\kk \equiv \Upsilon_\kk^\uparrow$ with
(\ref{UpsUpApp}) and approximating the chemical potential $\mu$ by
(\ref{eq:muInTriv}) (by simultaneous solution of the exact
$4\times 4$-theory self-consistency equations for $\Delta$ and $\mu$),
using the indicated values for $E_\mathrm{b}/E_\mathrm{F}$. The latter
correspond to $\ln(k_\mathrm{F} a_\mathrm{2D}) = 1.00$, $2.00$,
$3.00$, respectively (cf.\ Table~\ref{tab:scales}). Dashed curves show
the approximate expression (\ref{eq:approxHcrit}) that is expected to be
valid in the regime where $|\Delta| \ll \mu$.}
\end{figure}

In the regime of small $|\Delta|$, for which Fig.~\ref{fig:muCompare}(b)
is an illustration, the approximations Eqs.~(\ref{eq:muInTop}) and
(\ref{eq:muInTriv}) are accurate over the entire range of Zeeman
energies $h$, including the critical value $h_\mathrm{c}$ where both
expression yield coinciding values. Thus, from the condition that the
right-hand sides of (\ref{eq:muInTop}) and (\ref{eq:muInTriv}) are equal,
we can obtain an approximate expression for the phase boundary in
$h$--$\lambda$ space,
\begin{equation}\label{eq:approxHcrit}
\frac{h_\mathrm{c}}{E_\mathrm{F}} \approx \left| 1 - \frac{1}{2} \left(
\frac{2 m \lambda}{\hbar^2 k_\mathrm{F}} \right)^2 \right| \qquad (
|\Delta| \ll \mu) \quad .
\end{equation}
The result (\ref{eq:approxHcrit}) is consistent with the expectation that
$h_\mathrm{c}\approx |\mu_\mathrm{c}|$ for $|\Delta|\ll\mu$, which
follows straightforwardly from (\ref{eq:hCrit}), in conjunction with the
validity of the approximation (\ref{eq:muInTriv}). Figure~\ref{fig:phase_b}
shows the phase boundary calculated within the projected $2\times 2$
theory by solving the self-consistency condition (\ref{eq:practSC}) by
setting $\Upsilon_\kk \equiv \Upsilon_\kk^\uparrow$ with
(\ref{UpsUpApp}) and approximating $\mu/E_\mathrm{F}$ by
(\ref{eq:muInTriv}) while also enforcing the relation (\ref{eq:hCrit}). For
comparison, the approximation (\ref{eq:approxHcrit}) and exact results
obtained from the $4\times 4$ formalism are also included in these plots.
[The phase boundary found within the $2\times 2$-projected theory by
simultaneously self-consistent determination of $\Delta^{2\times
2}_\mathrm{c}$ and $\mu^{2\times 2}_\mathrm{c}$ differs only
imperceptibly from the more easily obtained $2\times 2$-theory curve
shown in Fig.~\ref{fig:phase_b} where $\mu_\mathrm{c}/E_\mathrm{F}$
is approximated by (\ref{eq:muInTriv}) in the calculation of the critical
Zeeman energy.] We restrict ourselves to showing the phase boundary
only for intermediate values of $2 m\lambda/(\hbar^2 k_\mathrm{F})$
where superfluidity is not expected to be destabilized by phase
separation~\cite{Zhou2011,Yang2012}.

Interestingly, the approximated $2\times 2$ approach turns out to predict
the phase boundary between topological and non-topological phases
correctly over a broader range of spin-orbit-coupling strengths than
na{\"\i}vely expected when considering the deviations between
$\Delta_\mathrm{c}$ and $\tilde\Delta^{2\times 2}_\mathrm{c}$ given in
Table~\ref{tab:scales}. This is the result of $h_\mathrm{c}$ being
generally dominated either by the value of $\mu_\mathrm{c}$ or that of
$\Delta_\mathrm{c}$. Although the $2\times 2$-projected theory
significantly underestimates $\Delta_\mathrm{c}$ for small $\lambda$,
$h_\mathrm{c}$ is dominated by the chemical potential in this parameter
range, in which the expression (\ref{eq:muInTriv}) for $\mu_\mathrm{c}$
is highly accurate. On the other hand, for large-enough $\lambda$ when
$\Delta_\mathrm{c}$ starts to become more important than
$\mu_\mathrm{c}$ for determining $h_\mathrm{c}$, the $2\times 2$
approach yields quite accurate values for $\Delta$. As a result, the
$h_\mathrm{c}(\lambda)$ dependence obtained within the projected
$2\times 2$ theory faithfully reproduces known qualitative features such
as the minimum at $2 m \lambda/(\hbar^2 k_\mathrm{F}) \gtrsim
1$~\cite{Yang2012}.

\subsection{Parametric dependences of the self-consistent
\textit{s}-wave pair potential}

\begin{figure}[t]
\begin{center}
\includegraphics[height=3.88cm]{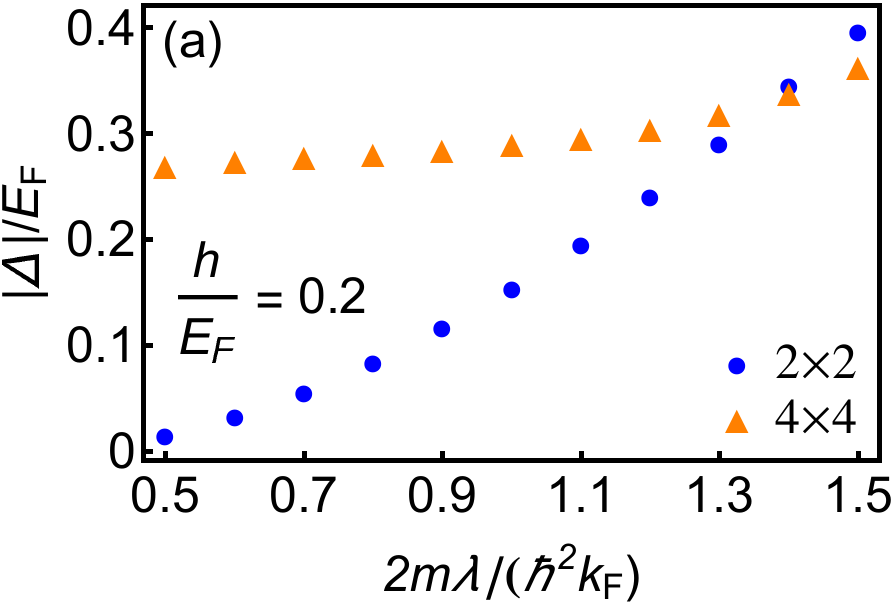}
\hspace{0.5cm}\includegraphics[height=3.88cm]{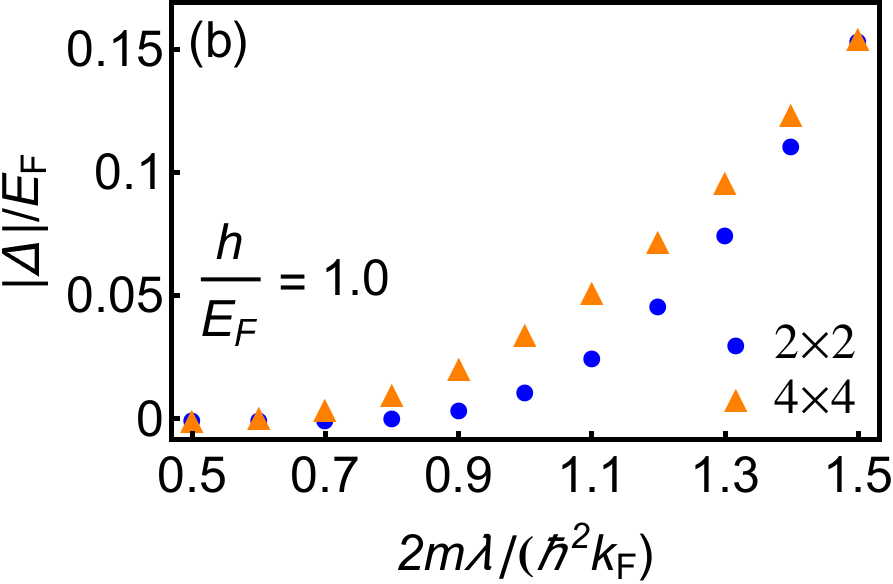}

\vspace{0.2cm}
\includegraphics[height=3.8cm]{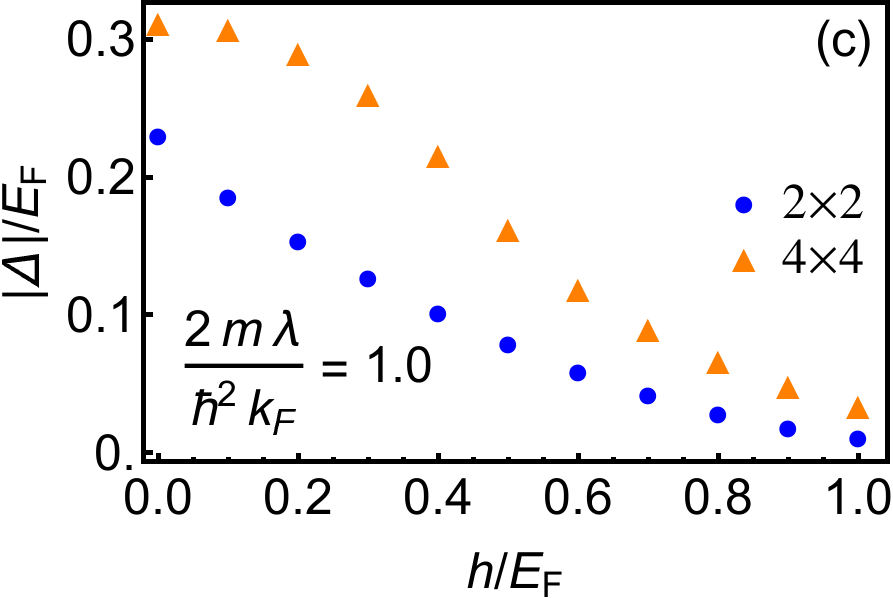}
\hspace{0.75cm}\includegraphics[height=3.8cm]{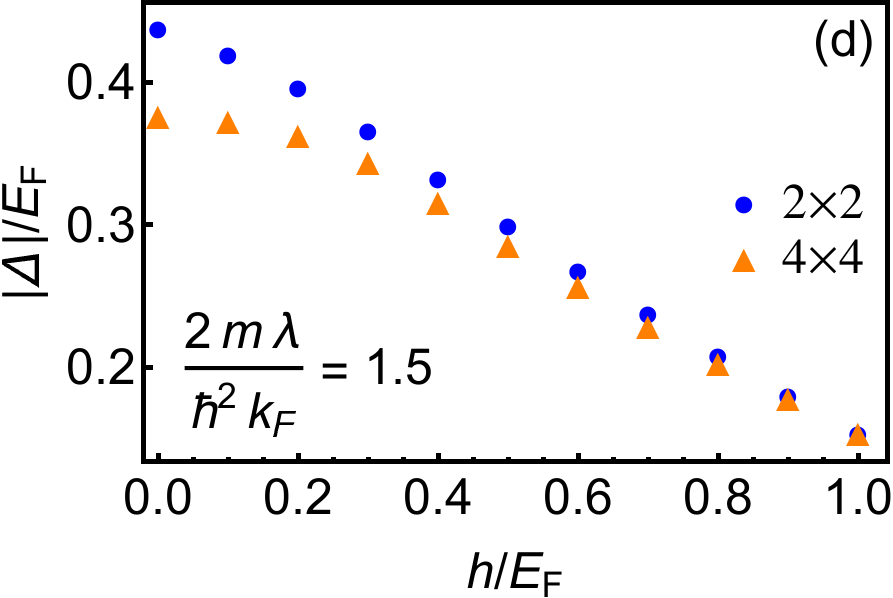}
\end{center}
\vspace{-0.5cm}
\caption{\label{fig:delta}%
Magnitude $|\Delta|$ of the \textit{s}-wave pair potential  obtained
self-consistently as a function of spin-orbit coupling strength $\lambda$
and Zeeman energy $h$. Data points labeled $2\times 2$ ($4\times 4$)
were calculated by solving the self-consistency equation
(\ref{eq:practSC}) for the \textit{s}-wave pair potential using
$\Upsilon_\kk \equiv \Upsilon_\kk^\uparrow$ with (\ref{UpsUpApp}) and
approximating the chemical potential $\mu$ by Eqs.~(\ref{eq:muInTop})
and (\ref{eq:muInTriv}) as appropriate (by simultaneous solution of the
exact $4\times 4$-theory self-consistency equations for $\Delta$ and
$\mu$), using $E_\mathrm{b}/E_\mathrm{F} = 0.0462$. The system is
in the non-topological [topological] superfluid phase for all data points
shown in panel (a) [(b)]. The critical Zeeman energy $h_\mathrm{c}$ is
equal to $0.507\, E_\mathrm{F}$ [$0.356\, E_\mathrm{F}$] for the
situation depicted in panel (c) [(d)].}
\end{figure}

The magnitude $|\Delta|$ of the \textit{s}-wave pair potential depends
intricately on the tunable system parameters $n$, $h$, $\lambda$, and
$E_\mathrm{b}$ through the self-consistency conditions
(\ref{eq:practSC}) and (\ref{eq:numbSelf}). As it turns out, the
dependence on the particle density $n$ is most conveniently absorbed
by using the Fermi wave vector $k_\mathrm{F}$ and Fermi energy
$E_\mathrm{F}$ as units for all other quantities to be measured in.
Figure~\ref{fig:delta} illustrates the $\lambda$ and $h$ dependence of
$|\Delta|$ and provides a comparison between results obtained within
the approximate $2\times 2$ approach and the exact $4\times 4$ theory.
[Both the simplified $2\times 2$-theory self-consistency routine where
$\mu_\mathrm{c}/E_\mathrm{F}$ is approximated by (\ref{eq:muInTriv})
and the simultaneously self-consistent determination of $\mu$ and
$|\Delta|$ within the $2\times 2$-projected approach yield practically
indistinguishable results for the parameters chosen in the Figure.] We
show numbers pertaining to fixed $E_\mathrm{b}/E_\mathrm{F} =
0.0462$, corresponding to $\ln(k_\mathrm{F} a_\mathrm{2D}) = 2$, to
enable direct comparison also with previous works~\cite{He2012a,
He2013} that give numerical results for $|\Delta|$ \textit{vs}.\
$h$~\footnote{The $|\Delta|$-\textit{vs}.-$\lambda$ dependence for a 2D
Fermi superfluid was explored before in Ref.~\cite{Yang2012}, albeit for
a situation where the spin polarization $(n_\uparrow - n_\downarrow)/n$
was held fixed instead of the Zeeman energy $h$. However, as can be
seen from Fig.~4 in that work, $h$ turns out to be effectively constant in
the range $2 m \lambda/(\hbar^2 k_\mathrm{F}) \gtrsim 0.5$ relevant for
our present study. Thus we can safely compare our results for the
$|\Delta|$-\textit{vs}.-$\lambda$ dependence, at the very least its
qualitative behavior, with that presented in Ref.~\cite{Yang2012}.}.

From the derivation of the main decoupling approximation
\eqref{eq:approx} of the projected approach, we may expect good
agreement between the $2\times 2$ and $4\times 4$ results when
$|\Delta| \ll h$, which is generally supported by the results reported in
Fig.~\ref{fig:delta}. It should be noted, though, that obtaining a small
$|\Delta|$ from the self-consistent $2\times 2$ theory is not sufficient to
guarantee this situation, as can be seen from Fig.~\ref{fig:delta}(a),
where  $|\Delta| / h >1$ according to the $4\times 4$ equations but the
accidental compensation of positive and negative parts of the summand
(as shown in Fig.~\ref{fig:gapIntComp}) results in small values for
$|\Delta|$ within the approximate $2\times 2$ theory. In such situations,
the projected $2\times 2$ approach typically tends to underestimate the
value of the self-consistent $|\Delta|$.

Figure~\ref{fig:delta}(a) [\ref{fig:delta}(b)] shows the $\lambda$
dependence of $|\Delta|$ for a situation where the system is in the
non-topological [topological] superfluid phase. The same qualitative
behavior of $|\Delta|$ increasing for increased $\lambda$ is exhibited in
both panels (a) and (b), for both the $2\times 2$ and $4\times 4$ data
points. However, the much weaker $|\Delta|$-\textit{vs.}-$\lambda$
dependence in the non-topological phase is not reproduced correctly by
the approximate $2\times 2$ formalism, whereas there is quite good
agreement with the exact results in the topological phase. This is
expected, as the projected $2\times 2$ theory should be more accurate
at larger $h$. The situation shown in our Fig.~\ref{fig:delta}(a)
corresponds reasonably closely to the case for which $|\Delta|$
\textit{vs}.\ $\lambda$ is plotted in Figs.~4(a) and 4(b) in
Ref.~\cite{Yang2012} (they have a larger $E_\mathrm{b}/E_\mathrm{F}$
and smaller $h/E_\mathrm{F}$), and there is excellent qualitative
agreement between their results and ours.

The exact results for the $|\Delta|$-\textit{vs.}-$h$ dependence given in
Fig.~\ref{fig:delta}(c) agree with those available from
Refs.~\cite{He2012a,He2013}. For small $h$, deviations between the
values calculated within the projected $2\times 2$ formalism and the
exact $4\times 4$ theory are significant, and even the qualitative
behavior exhibited by the respective $|\Delta|$-\textit{vs}.-$h$
dependences is seen to be quite different. However, the agreement
becomes quite good in the topological regime realized for $h >
h_\mathrm{c} = 0.507\, E_\mathrm{F}$. In contrast, the projected
$2\times 2$ theory is seen to become overall very accurate, even in the
non-topological phase, for the larger value of $\lambda$ for which
results are given in Fig.~\ref{fig:delta}(d). Thus, as already indicated by
the numbers in Table~\ref{tab:scales}, the effective $2\times 2$ theory
yields quantitatively satisfactory results for sufficiently large values of $2
m\lambda/(\hbar^2 k_\mathrm{F})$.

\section{Conclusions and outlook}\label{sec:concl}

We have derived an accurate effective theoretical description for
superfluidity in 2D Fermi gases with broken spin-rotational invariance
due to the presence of spin-orbit coupling and Zeeman spin splitting.
Starting from the usually applied self-consistent Bogoliubov-de~Gennes
(BdG) mean-field theory for \textit{s}-wave pairing in four-dimensional
Nambu space [Eq.~(\ref{eq:bdg}) with (\ref{eq:origBdG}) and
(\ref{eq:practSC})], we performed a Feshbach projection onto subspaces
associated with fixed spin-$\sigma$ degrees of freedom
[Eqs.~(\ref{eq:Feshbach})]. Using also the approximations given in
Eqs.~(\ref{eq:approx}) that are informed by inspection of limiting
behaviors in the BdG quasiparticle dispersions and ignoring terms of
$\mathcal{O}(|\Delta|/h)$, we succeeded in fully decoupling the original
$4\times 4$ BdG equation (\ref{eq:bdg}) into two $2\times 2$ BdG
equations; one for each spin projection [Eq.~(\ref{eq:2bandBdG}) with
(\ref{eq:BdGapprox})].

Our subsequent investigations focusing on uniform systems at fixed total
particle density have demonstrated that the effective two-band
descriptions for individual spin subspaces provide a useful theoretical
framework for studying the unusual physical properties of this superfluid,
including topological effects. In particular, we found that the effective
theory faithfully reproduces the relevant physical aspects of the
Bogoliubov-quasiparticle dispersion with the chiral-\textit{p}-wave-like
gap [see Figs.~\ref{fig:dispersions}(b) and \ref{fig:moreDisp}] and the
occupation-number distribution in reciprocal space (see
Fig.~\ref{fig:densComp}). For both the dispersions and reciprocal-space
density distributions, the projected-$2\times 2$-theory's accuracy is
excellent in the topological regime but generally very good even within a
finite range on the non-topological side of the transition. As the Zeeman
spin-splitting energy $h$ decreases, so does the accuracy of the
projected $2\times 2$ approach. This is most apparent in the
comparison of the chemical potentials self-consistently obtained within
the exact $4\times 4$ and approximate $2\times 2$ approaches,
respectively, shown in Fig.~\ref{fig:muCompare}. Based on the
observation of Fermi-surface-like features in the reciprocal-space
occupation-number distribution [Figs.~\ref{fig:densComp}(c) and 
\ref{fig:densComp}(f)], we derived analytical formulae for the chemical
potential [Eqs.~(\ref{eq:muInTop}) and (\ref{eq:muInTriv})] that agree
very well with the exact results (see Fig.~\ref{fig:muCompare}).

The self-consistency condition for the \textit{s}-wave pair potential within
the $2\times 2$ theory turns out to be given entirely in terms of quantities
relating to the spin-$\uparrow$ states [Eq.~(\ref{eq:practSC}) where
$\Upsilon_\kk$ as defined in (\ref{eq:Upsilon}) is replaced by
$\Upsilon_\kk^\uparrow$ from (\ref{UpsUpApp})]. We devise two routines
for achieving full self-consistency within the $2\times 2$-projected
theory. One is based on the simultaneous solution of the
self-consistency conditions (\ref{eq:practSC}) and (\ref{eq:numbSelf})
using $2\times 2$-theory results as input: $n_{\kk\uparrow}\equiv n_{\kk
\uparrow}^\uparrow$, $n_{\kk\downarrow}\equiv n_{\kk
\downarrow}^\downarrow + n_{\kk\downarrow}^\uparrow$, and
$\Upsilon_\kk\equiv \Upsilon_\kk^\uparrow$ with relevant expressions
given in Eqs.~(\ref{eq:diagSpinDen}), (\ref{eq:upSpinFromDown}), and
(\ref{UpsUpApp}). The other, simpler routine solves the self-consistency
condition (\ref{eq:practSC}) using $\Upsilon_\kk\equiv
\Upsilon_\kk^\uparrow$ as given in Eq.~(\ref{eq:Upsilon}) and with
the chemical potential approximated by the analytical expressions from
Eqs.~(\ref{eq:muInTop}) and (\ref{eq:muInTriv}). For the parameter
ranges explored in this work, both routines yield practically
indistinguishable results for $|\Delta|$, thus making it possible to adopt
the simpler routine for further exploration of the physical ramifications of
the $2\times 2$-projected theory. Overall, the combination of the
projected two-band description for spin-$\uparrow$ states with the
analytical formulae for the chemical potential is seen to provide a reliable
theoretical description of the system, with impressive quantitative
agreement achieved in the limit of sufficiently large, but entirely realistic,
values of the Zeeman splitting and spin-orbit coupling (see
Figs.~\ref{fig:phase_b}, \ref{fig:delta} and Table~\ref{tab:scales}). 

The ability to utilize an effective two-band ($2\times 2$) theory for
describing superfluidity in the 2D spin-split Fermi gas opens up the
opportunity to explore in greater detail suggested analogies with
chiral-\textit{p}-wave pairing~\cite{Kallin2016}. In particular, based on the
demonstrated accuracy of the projected $2\times 2$ approach for the
case of uniform systems, we expect this formalism to also be effective
for describing situations with non-uniform order-parameter
configurations~\cite{Grosfeld2011,Zou2016,Smith2016} or in the
presence of disorder~\cite{Shitade2015,Goertzen2017}. These
scenarios are interesting because they offer possibilities to create and
manipulate exotic Majorana excitations spatially~\cite{Gurarie2007,
Liu2012b} or temporally~\cite{Zou2016}. Future work will address in
detail the question of applicability of the $2\times 2$ approach in such
instances and, as appropriate, apply it to inform the design of basic
building blocks for fault-tolerant quantum information
processing~\cite{DasSarma2015,Beenakker2016}.

\section*{Acknowledgements}
The authors gratefully acknowledge useful discussions with Philip
M.~R.\ Brydon and Ana Maria Rey, as well as Kadin Thompson's technical
help with numerical calculations.

\paragraph{Funding information}
This work was partially supported by the Marsden Fund of New Zealand
(contract no.\ MAU1604), from government funding managed by the
Royal Society Te Ap$\bar{\mathrm a}$rangi .

%
%


\nolinenumbers

\end{document}